\documentclass[12pt, draftclsnofoot, onecolumn]{IEEEtran}

\usepackage{cite}
\usepackage{mathtools}
\usepackage{hyperref}
\usepackage{graphicx}
\usepackage{enumitem}
\usepackage{tikz}
\usepackage{bm}
\usepackage{amsmath}
\usepackage{algorithm}
\usepackage{algorithmic}
\usepackage{amsmath,amssymb,amsthm}
\usepackage{array}
\usepackage{relsize}
\usepackage{subcaption}
\usepackage{mathtools, nccmath}
\usepackage[labelformat=simple]{subcaption}

\newcolumntype{P}[1]{>{\centering\arraybackslash}p{#1}}
\DeclareMathOperator{\C}{\mathbb{C}}
\DeclareMathOperator{\E}{\mathbb{E}}
\DeclareMathOperator{\R}{\mathbb{R}}
\DeclareMathOperator{\N}{\mathcal{N}}
\DeclareMathOperator{\LA}{\mathcal{L}}
\DeclareMathOperator{\CG}{\mathcal{C}}
\DeclareMathOperator{\BO}{\mathcal{O}}
\DeclareMathOperator*{\argmax}{arg\,max}
\DeclareMathOperator*{\argmin}{arg\,min}
\DeclarePairedDelimiter\abs{\lvert}{\rvert}
\DeclareCaptionSubType * [alph]{table}
\newcommand{\A}{\mathbf{A}}
\newcommand{\B}{\mathbf{B}}
\newcommand{\h}{\mathbf{H}}
\newcommand{\f}{\mathbf{F}}

\newcommand{\I}{\mathbf{I}}
\newcommand{\z}{\mathbf{Z}}

\newcommand{\U}{\mathbf{U}}
\newcommand{\V}{\mathbf{V}}

\newcommand{\fs}{\mathbf{f}}
\newcommand{\rt}{\mathbf{R}}
\newcommand{\D}{\mathbf{D}}
\newcommand{\as}{\mathbf{a}}
\newcommand{\bs}{\mathbf{b}}
\newcommand{\q}{\mathbf{Q}}
\newcommand{\js}{{\rm j}}
\newcommand{\cn}{\mathsf{c}}
\newcommand{\gs}{\mathbf{g}}
\newcommand{\dk}{\mathbf{d}}
\newcommand{\rs}{\mathbf{r}}
\newcommand{\vs}{\mathbf{v}}
\newcommand{\us}{\mathbf{u}}

\newcommand{\hs}{\mathbf{h}}

\newcommand{\sn}{\mathsf{s}}
\newcommand{\xs}{\mathbf{x}}
\newcommand{\xsn}{\mathsf{x}}
\newcommand{\xn}{\bm{\mathsf{x}}}
\newcommand{\ysn}{\mathsf{y}}
\newcommand{\ys}{\bm{\mathsf{y}}}
\newcommand{\s}{\bm{\mathsf{s}}}
\newcommand{\ws}{\bm{\mathsf{w}}}
\newcommand{\wsn}{\mathsf{w}}
\newcommand{\zs}{\mathbf{z}}
\newcommand{\zn}{\bm{\mathsf{z}}}
\newcommand{\hr}{\mathsf{H}}
\newcommand{\tr}{\mathsf{T}}
\newcommand{\1}{\mathbf{1}}
\newcommand{\0}{\mathbf{0}}
\newcommand{\bc}{\begin{center}}
\newcommand{\ec}{\end{center}}
\newcommand{\norm}[1]{\left\lVert#1\right\rVert}
\newcommand{\ds}{\displaystyle}
\newcommand{\uprightsubscript}[1]{_{\textnormal{#1}}}
\begingroup\lccode`~=`?\lowercase{\endgroup\let~}\uprightsubscript
\AtBeginDocument{\mathcode`?="8000 }
\newcommand{\<}[1]{^{\textnormal{#1}}}
\newcommand{\tn}[1]{\textnormal{#1}}

\theoremstyle{remark}
\newtheorem*{remark}{Remark}
\definecolor{dg}{RGB}{0,0,0}

\begin{document}
\title{Joint Active and Passive Beamforming Design for IRS-Assisted Multi-User MIMO Systems: A VAMP-Based Approach} 
\author{\IEEEauthorblockN{Haseeb Ur Rehman, Faouzi Bellili, \textit{Member, IEEE}, Amine Mezghani, \textit{Member, IEEE}, and Ekram Hossain, \textit{Fellow, IEEE}} \thanks{The authors are with the Department of Electrical and Computer Engineering at the University of Manitoba, Canada (emails: urrehmah@myumanitoba.ca, \{Faouzi.Bellili, Amine.Mezghani, Ekram.Hossain\}@umanitoba.ca).}}
\date{}
\maketitle
\vspace{-40pt}
\textcolor{dg}{
\begin{abstract}
This paper tackles the problem of joint active and passive beamforming optimization for an intelligent reflective surface (IRS)-assisted multi-user downlink multiple-input multiple-output (MIMO) communication system under both ideal and practical IRS phase shifts. We aim to maximize the spectral efficiency of the users by minimizing the sum mean square error (MSE) of the users' received symbols. For this, a joint non-convex optimization problem is formulated under the sum minimum mean square error (MMSE) criterion. Alternating minimization is used to break the original joint optimization problem into the separate optimization of the active precoding matrix for the base station (BS) and the matrix of phase shifts for the IRS. While the MMSE active precoder is obtained in closed-form, the IRS phase shifts are optimized iteratively using a modified version (developed in this paper) of the vector approximate message passing (VAMP) algorithm. Moreover, the underlying joint optimization problem is solved under two different models for the IRS phase shifts, namely by assuming $i)$
a unimodular (i.e., ideal) constraint on the reflection coefficients and $ii)$ a more practical  reflection elements termination by a variable reactive load (which inherently introduces the phase-dependent amplitude attenuation in the IRS phase shifts). Simulation results are presented to illustrate the performance of the proposed method under both perfect and imperfect channel state information (CSI) and to show the effect of the practical constraint on the system throughput. The results validate the superiority of the proposed method over the state-of-the-art techniques both in terms of throughput and  computational complexity. 
\end{abstract}
}

\section{Introduction}

\subsection{Background}
The need for higher data rates in wireless communication is soaring. This calls for innovative and economically viable communication technologies that can keep up with the increasing network capacity requirement.
Massive multiple-input multiple-output (MIMO) technology  can fulfill the network capacity requirement for beyond fifth-generation (B5G) wireless networks \cite{rusekmimo,larssonmimo,marzmimo}. The basic idea of massive MIMO is to equip the base stations (BSs) with tens (if not hundreds) of antenna elements so as to simultaneously serve multiple mobile devices using the same time/frequency resources. Despite the many advantages of massive MIMO, its practical large-scale deployment is hindered by the associated high hardware cost and energy consumption~\cite{buzzimimo,zhangmimo}. Moreover, although millimeter wave (mmWave) communication benefits from massive MIMO due to a symbiotic convergence of technologies, its practical use is still limited by the less penetrative propagation characteristic of mmWave signals in presence of blockages between the BS and the mobile device. \cite{tanmimo}.

One promising technology that has been introduced recently is intelligent reflective surfaces (IRSs), also called reconfigurable intelligent surfaces (RISs)~\cite{liasirs,huirs}. IRS is composed of a planar metasurface consisting of a large number of passive reflective elements. Moreover, IRS does not require a power amplifier for transmission which makes it an energy-efficient technology. This allows the IRS to passively alter the signal propagation by reconfiguring the phases of its reflective elements through a controller attached to the surface~\cite{panirs}. Therefore, IRSs can be utilized to perform passive beamforming. Passive beamforming refers to changing the IRS phases without actively powering the IRS antenna elements as opposed to active beamforming at the BS so as to improve the received power while reducing the interference for unintended users, thereby enhancing the overall throughput of the network~\cite{wuirs}. Practically, IRS deployment requires a large number of cost-effective phase shifters (PSs) on a surface that can be easily integrated into a traditional wireless network~\cite{huangirs}.
Due to the aforementioned reasons, IRS-assisted communication has gained substantial research interest in the wireless  research community over the recent few years \cite{wuirs,huangirs,qwuirs,zhouirs,yangirs,yuirs,wangirs,abeyirs,zhengirs}. 

\textcolor{dg}{In \cite{yangirs}, a multi-user multiple-input single-output (MISO) wireless system assisted by a single IRS in the downlink configuration is studied. The authors present a deep reinforcement learning (DRL)-based solution to jointly optimize the IRS phase shifts and the BS precoding under different quality of service constraints. In \cite{zhengirs}, authors tackle the problem of estimating the cascaded BS-IRS-user channels for an IRS-assisted multi-user MISO system. The author propose a pilot-based solution and improve its efficiency by exploiting the fact that all users share the same BS-IRS channel. The same problem is solved by utilizing the deep residual learning framework in \cite{liuirs}.} In \cite{liirs}, an IRS-assisted multi-cluster MISO system serving multiple users is considered wherein the authors seek to minimize the transmit power under a minimum signal-to-interference-plus-noise ratio (SINR) constraint by jointly optimizing the IRS phase shifts and the transmit precoder. They tackle the underlying problem through alternating direction method of multipliers (ADMM).
An IRS-aided MISO and MIMO system with discrete phase shifts for IRS elements is also discussed in \cite{qwuirs}. The authors formulate the problem of minimizing the transmit power under minimum SINR constraint and jointly optimize the transmit precoding and the IRS phase shifts in a mixed-integer non-linear programming framework. In \cite{abeyirs}, a relatively more practical model for IRS reflection coefficients is considered, and then a penalty-based algorithm is used to optimize the phase matrix. 

The vast majority of the existing work considers a MISO wireless system assisted by a single or multiple IRSs serving a single user~\cite{zhouirs,wangirs,abeyirs}. So far, limited research has been conducted on IRS-aided multi-user MIMO systems. Moreover, IRS reflection coefficients are often modeled as ideal phase shifters and a realistic approach towards modeling reflection coefficients has rarely been investigated. In fact, most of the existing methods are limited to a single-phase shifter model, unimodular phase shifts being the most common one, and hence they are not resilient to the various hardware impairments of the IRS reflection elements~ \cite{zhouirs,qwuirs,wangirs,wuirs,abeyirs}.

\subsection{Contributions}

In this paper, we consider a multi-user IRS-assisted single-cell downlink MIMO system with a single IRS. The IRS is equipped with a large number of passive phase shifters that aid the BS to serve a small number of users. We propose a robust solution for the problem of jointly optimizing the active and passive beamforming tasks under different models for the IRS reflection coefficients. The main contributions embodied by this paper are as follows:
\begin{itemize}
\item We solve the problem of maximizing the spectral efficiency of the users by jointly optimizing the transmit precoding matrix at the BS and the reflection coefficients at the IRS. To that end, we first formulate the joint optimization problem under the sum MMSE criterion in order to minimize the MSE of the received symbols for all users at the same time.

\item To solve the underlying joint optimization problem, we first split it using alternate optimization ~\cite{dimitrinl} into two easier sub-optimization tasks of the active precoder at the BS and the reflection coefficients at the IRS.
The precoding sub-optimization problem is similar to the MMSE transmit precoder optimization for a traditional MIMO system, which can be solved in closed-form through Lagrange optimization.
 
\item We modify and extend the existing VAMP algorithm \cite{ranganvamp} and propose a robust technique to find locally optimal reflective coefficients for the IRS under multiple constraints. Precisely, we find the sub-optimal phase matrix under two different models for the reflection coefficients: $i)$ Under the unimodular constraint on the IRS reflection coefficients and $ii)$ under a practical constraint, where each IRS element is terminated by a tunable simple reactive load.
\textcolor{dg}{
\item We discuss the convergence and provide the order of complexity of the proposed solution. We present various numerical results to compare the proposed solution with the semi-definite relaxation (SDR) plus MMSE-based IRS beamforming and precoding optimization approach \cite{wusdrc,wuirs}, an ADMM-based solution, and a standalone massive MIMO system using MMSE precoder. The results show that, the proposed solution: $i)$ outperforms both the SDR-based and the ADMM-based solutions in terms of throughput while using the same resources and being less computationally demanding, and $ii)$ achieves higher throughput than a traditional massive MIMO system while using a significantly smaller number of transmit antennas in typical propagation scenarios. We illustrate the effect of practical phase shifts on the system throughout. We also show the robustness of the proposed solution by assessing its performance under imperfect CSI.}
\end{itemize} 

\subsection{Paper Organization and Notations}

The rest of the paper is organized as follows: the system model along with the problem formulation for jointly optimizing the active precoder and the reflection coefficients are discussed in Section \ref{sec:system-model}. Section \ref{sec:vamp} briefly introduces the VAMP algorithm and then extends it to solve optimization problems. In Section \ref{sec:joint-phase-precod}, we solve the optimization problem at hand using the proposed extended version of VAMP. In Section \ref{sec:reactive-loading}, we further solve the underlying optimization problem under the ``simple reactive loading" constraint on the IRS reflection elements. Exhaustive numerical results are shown in Section \ref{sec:numeric-res}. Finally, Section \ref{sec:complexity-con-opt} provides an analysis on the convergence and computational complexity of the proposed solution.

\vspace{0.1cm}
\noindent
\textbf{Notations}: Lowercase letters (e.g., $r$) denote scalar variables. The uppercase italic letters (e.g., $N$) represent scalar constants. Vectors are denoted by small boldface letters (e.g., $\zs$) and the $k$-th element of $\zs$ is denoted as $z_k$. Exponent on a vector (e.g., $\zs^n$) denotes component-wise exponent on every element of the vector. Capital boldface letters (e.g., $\A$) are used to denote matrices, while $a_{ik}$ and $\bm{a}_i$ stand, respectively, for the $(i,k)$-th entry and the $i$-th column of $\A$. $\C^{ M \times N}$ denotes the set of matrices of size $M \times N$ with complex elements and $\A^{-k}$ means $\left(\A^{-1}\right)^k$. $\tn{Rank}(\A)$ and $\tn{Tr}(\A)$, return, respectively, the rank and the trace of any matrix $\A$. We also use $\|.\|?2, \ \|.\|?F, \ (.)^*, \ (.)^\tr, \ (.)^\hr$ to denote the $\mathcal{L}?2$ norm, Frobenius norm, the conjugate, the transpose, and the conjugate transpose operators, respectively. The operator $<.>$ returns the empirical average of all the elements/entries of any vector or matrix. Moreover, $\tn{vec}(.)$ and $\tn{unvec}(.)$ denote vectorization of a matrix and unvectorization of a vector back to its original matrix form, respectively. $\tn{Diag}(.)$ operates on a vector and generates a diagonal matrix by placing that vector in the diagonal whereas $\tn{diag}(.)$ operates on a matrix and returns its main diagonal in a vector. The statistical expectation is denoted as $\E\lbrace.\rbrace$.  A random vector with complex normal distribution is represented by $\xs \sim \CG\N(\xs;\us,\rt)$, where $\us$ and $\rt$ denote its mean and covariance matrix, respectively. The imaginary unit is represented by $\js=\sqrt{-1}$ and the $\angle (.)$ operator returns the angle of its complex argument. The proportional relationship between any two entities (functions or variables) is represented by $\varpropto$ operator. Lastly, the operators $\otimes$, $\odot$ and $\ast$ denote the Kronecker, the Hadamard and the column-wise Khatri-Rao products, respectively. 

\section{System Model, Assumptions, and Problem Formulation}
\label{sec:system-model}

Consider a BS that is equipped with $N$ antenna elements serving $M \ (M<N)$ single-antenna users in the downlink. The BS is assisted by an IRS which has $K \ (K > M)$ reflective elements.
For each $m$-th user, we have a direct link to the BS expressed by a channel vector $\hs_{\tn{b-u},m} \in \C^{N}$. The channel of the surface-user $m$ link is denoted by $\hs_{\tn{s-u},m} \in \C^{K}$. 
\textcolor{dg}{Let $\h?{b-s} \in \C^{K \times N}$ denote the channel matrix of the MIMO IRS-BS link with $\tn{Rank}(\h?{b-s})\geq M$.} The signal received at the IRS is phase-shifted by a diagonal matrix
$\mathbf{\Upsilon}=\tn{Diag}(\bm\upsilon) \in \C^{K \times K}$, where $\bm\upsilon \in \C^K$ is the phase-shift vector
having unimodular elements, i.e., $\abs*{\upsilon_k}=1$ for $k=1,\cdots,K$. In other words, for each reflection element, we have $\upsilon_k=e^{\js\theta_k}$ for some phase shift $\theta_k\in[0,2\pi]$. The received signal for user $m$ can be expressed as follows:
\begin{equation}
 \ysn_m=\alpha\left(\hs_{\tn{s-u},m}^\hr{\mathbf{\Upsilon}}\h?{b-s}\ds\sum_{m'=1}^M\fs_{m'}\sn_{m'}  + \hs_{\tn{b-u},m}^\hr\ds\sum_{m'=1}^M\fs_{m'}\sn_{m'} +\ \wsn\right),~~m=1, \cdots, M,
\end{equation}
where $\sn_m \sim \CG\N(s;0,1)$ is the unknown transmit symbol, $\wsn \sim \CG\N(w;0,\sigma?w^2)$ denotes additive white Gaussian noise (AWGN), and $\alpha \in \R$ refers to the receiver scaling which is a common practice in precoding optimization literature \cite{nossek2005,jeddaprecoding}. Here, $\fs_m \in \C^{N}$ for $m=1,\cdots, M$ are the precoding vectors that are used for power allocation and beamforming purposes.
Let $\f=[\fs_1, \fs_2, \cdots, \fs_M]$ be the precoding matrix and let $P$ denote the total transmit power. By denoting $\s=[\sn_1,\sn_2,\cdots,\sn_M]^\tr$, it follows that $\E\big\{\|\f\s\|^2\big\} = P$.
Let  $\h?{b-u} = \left[\hs_{\tn{b-u},1}, \hs_{\tn{b-u},2}, \cdots, \hs_{\tn{b-u},M}\right]$ and $\h?{s-u} = [\hs_{\tn{s-u},1}, \hs_{\tn{s-u},2}, \cdots, \hs_{\tn{s-u},M}]$. Then, by stacking all the users' signals in one vector $\ys=[\ysn_1, \ysn_2, \cdots, \ysn_M]^\tr$, we can express the input-output relationship of the multi-user MIMO system as:
\begin{equation}
\ys=\alpha\left(\underbrace{\h?{s-u}^\hr{\mathbf\Upsilon}\h?{b-s}\f\s}?{Users-IRS-BS}  + \underbrace{\h?{b-u}^\hr\f\s}?{Users-BS} +\ \ws\right).
\end{equation}
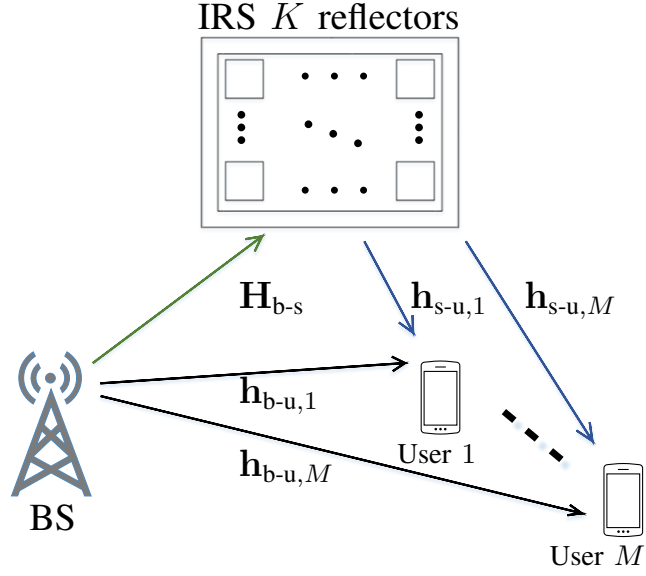
\begin{figure}
\vspace{-10pt}
\bc
\scalebox{1.2}{
\begin{picture}(250,180)
\put(0,0){\includegraphics[scale=0.45]{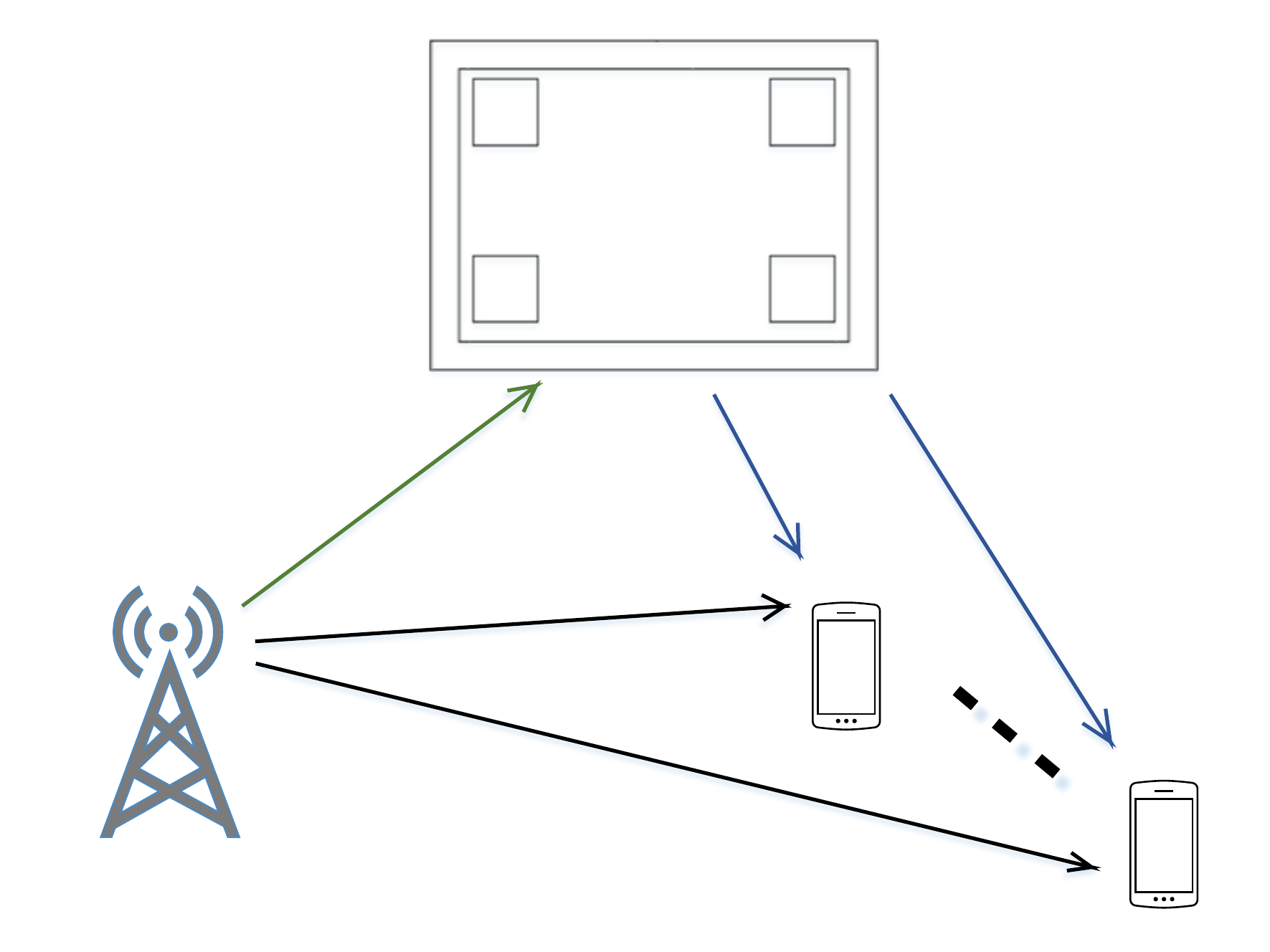}}
\put(24,10){BS}
\put(77,168){IRS $K$ reflectors}
\put(140,31){\footnotesize{User $1$}}
\put(188,-1){\footnotesize{User $M$}}
\put(90,50){$\hs_{\tn{b-u},1}$}
\put(90,27){$\hs_{\tn{b-u},M}$}
\put(144,80){$\hs_{\tn{s-u},1}$}
\put(180,80){$\hs_{\tn{s-u},M}$}
\put(90,80){$\h?{b-s}$}
\put(88,132){{\huge $\vdots$}}
\put(144,132){{\huge $\vdots$}}
\put(108,148){{\huge $\cdots$}}
\put(108,112){{\huge $\cdots$}}
\put(108,130){{\huge $\ddots$}}
\end{picture}
}
\caption{IRS-assisted multi-user MIMO system.}
\ec
\vspace{-10pt}
\end{figure}
The overall effective channel matrix for all users is thus given by:
\begin{equation}
\h^\hr=\h?{s-u}^\hr{\mathbf\Upsilon}\h?{b-s} +\h?{b-u}^\hr.
\end{equation}
We aim to minimize the received symbol error of each user under the MMSE criterion, which consequently maximizes the user SINR. A lower bound on the spectral efficiency for user $m$ can be expressed in terms of the MMSE of its received symbol \cite{heathmimo} as follows:
\begin{equation}
C_m\<{MMSE}=\log_2\left(\frac{1}{\tn{MMSE}_m}\right).
\end{equation}
The MSE of the received symbol for user $m$ is given by $\E_{\ysn_m,\sn_m}\left\lbrace\abs*{\ysn_m-\sn_m}^2\right\rbrace$, and for $M$ users, the sum symbol MSE can be written as:
\begin{equation}
\ds\sum_{m=1}^M\E_{\ysn_m,\sn_m}\left\lbrace\abs*{\ysn_m-\sn_m}^2\right\rbrace \ = \ \E_{\ys,\s}\left\lbrace\norm{\ys-\s}_2^2\right\rbrace.
\end{equation}
\vspace{-10pt}
Thus, our problem under the MMSE criterion can be formulated as follows:
\begin{subequations}
\label{eqn:mmse-opt}
\begin{align}
\label{eqn:mmse-opta}
\ds\argmin_{\alpha, \f, {\mathbf\Upsilon}} \quad & \E_{\ys,\s}\left\lbrace\|\ys-\s\|_2^2\right\rbrace,\\
\label{eqn:mmse-optb}
\tn{subject to} \quad & \E_{\s}\left\lbrace\|\f\s\|_2^2\right\rbrace=P,\\
& \upsilon_{ik}=0, \quad i \neq k,\\
& |\upsilon_{ii}|=1, \quad i=1,2, \cdots, K.
\end{align}
\end{subequations}
\vspace{-25pt}
\textcolor{dg}{
\begin{remark}
The objective function in \eqref{eqn:mmse-opta} leads to some fairness among the users by ensuring that the MSE is minimized for each user. The lower bound on sum-spectral-efficiency of $M$ users can be expressed in terms of the MMSE of the users' received symbols \cite{heathmimo} as follows:
\begin{equation}
\widehat{C} \ = \ \ds\sum_{m=1}^M\log_2\left(\frac{1}{\tn{MMSE}_m}\right) \ = \ \log_2\left(\ds\prod_{m=1}^M\frac{1}{\tn{MMSE}_m}\right).
\end{equation}
In other words, maximizing the sum-spectral-efficiency is equivalent to minimizing the product MSE of all users. This can be achieved by minimizing the MSE of the user with the strongest channel, thereby leading to a very unfair solution. On the other hand, aiming for complete fairness results in a very inefficient allocation of resources when it comes to the overall system throughput. In this respect, the sum MMSE criterion is a good balance between the two extremes. Since our aim is to maximize the spectral efficiency of each user rather than the sum-spectral-efficiency, the MMSE criterion is a good fit for our problem formulation. 
\end{remark}}
\vspace{-5pt}
The expectation involved in \eqref{eqn:mmse-opta} and \eqref{eqn:mmse-optb} is taken with respect to (w.r.t.) the random vectors $\s$ and $\ws$. Explicitly writing the objective function in \eqref{eqn:mmse-opta} leads to:
\vspace{-10pt}
\textcolor{dg}{
\begin{multline}
\E_{\ws,\s}\big\{\tn{Tr}\left(\alpha^2\s^\hr\f^\hr\h\h^\hr\f\s-\alpha\s^\hr\f^\hr\h\s-\alpha\s^\hr\h^\hr\f\s+\s\s^\hr\right)\\
+\tn{Tr}\left(\alpha^2\s^\hr\f^\hr\h\ws+\alpha^2\ws^\hr\h^\hr\f\s-\alpha\ws^\hr\s-\alpha\s^\hr\ws+\alpha^2\ws^\hr\ws\right)\big\},
\end{multline}}
\vspace{-5pt}
thereby resulting in the following optimization problem.
\begin{subequations}
\label{eqn:exp-mmse-opt}
\begin{align}
\label{eqn:exp-mmse-opta}
\ds\argmin_{\alpha, \f, {\mathbf\Upsilon}} \quad & \norm{\alpha\h?{s-u}^\hr{\bm\Upsilon}\h?{b-s}\f-(\I_M-\alpha\h?{b-u}^\hr\f)}?F^2 + M\alpha^2\sigma?w^2,\\
\label{eqn:exp-mmse-optb}
\tn{s.t.} \quad & \|\f\|?F^2=P,\\
\label{eqn:exp-mmse-optc}
& \upsilon_{ik}=0, \quad i \neq k,\\
\label{eqn:exp-mmse-optd}
& |\upsilon_{ii}|=1, \quad i=1,2, \cdots, K.
\end{align}
\end{subequations}
The optimization problem in \eqref{eqn:exp-mmse-opt} is non-convex optimization problem due to the unimodular constraint\footnote{Later, we will solve the same problem under another constraint on the reflection coefficients.} on the IRS phase shifts in \eqref{eqn:exp-mmse-optd}.
\textcolor{dg}{
VAMP is a low-complexity algorithm which is designed to solve optimization problems with a linear objective function and non-linear constraints \cite{ranganvamp}.
VAMP has a modular structure that makes it possible to decouple the constraints from the objective function. Therefore, the same objective function can be minimized under different constraints by modifying the VAMP module that satisfies the constraint (simple scalar functions). VAMP automatically updates the stepsize at a per-iteration basis that leads to a faster convergence compared to other iterative algorithms (e.g., ADMM) \cite{manvamp}. This favorable property makes VAMP tuning-free. 
The performance of VAMP can be theoretically predicted to establish optimality through the \mbox{statistical state evolution framework \cite{barbieroptimal,ranganvamp}. This is, however left to future work.}}
\vspace{-10pt}
\section{Modified VAMP Algorithm for Constrained Optimization}

\label{sec:vamp}
Recently, message passing algorithms~\cite{ranganvamp, montamp,bishop2006} have gained attention in estimation theory because of their high performance and fast convergence. Vector approximate message passing (VAMP)~\cite{ranganvamp}, in particular, is a  low-complexity algorithm that solves quadratic loss optimization of recovering a vector from noisy linear measurements.
In this section, we briefly discuss the standard max-sum VAMP algorithm and we further modify it to solve the constrained optimization problem at hand.
\vspace{-10pt}
\subsection{Background on Max-Sum VAMP}

Approximate message passing (AMP)-based computational techniques have gained a lot of attention since their introduction within the compressed sensing framework \cite{montamp}. To be precise, AMP solves the standard linear regression problem of recovering a vector $\xs \in \C^N$ from noisy linear observations:
\vspace{-5pt}
\begin{equation}
\zn=\A\xs+\ws,
\end{equation}
where $\A\in\C^{M \times N}$ (with $M \ll N$) is called sensing matrix and $\ws\sim\CG\N(\ws;0,\gamma?w^{-1}\I_M)$, with $\gamma?w > 0$, so that $p_{\zn|\xn}(\zs|\xs)=\CG\N(\zs;\A\xs,\gamma?w^{-1}\I_M)$.
Interestingly, the performance of AMP under independent and identically distributed (i.i.d.) Gaussian sensing matrices, $\A$, can be rigorously tracked through scalar state evolution (SE) equations \cite{montamp2}. One major drawback of AMP, however, is that it often diverges if the sensing matrix, $\A$, is ill-conditioned or has a non-zero mean. To circumvent this problem, vector AMP (VAMP) algorithm was proposed and rigorously analyzed through SE equations in \cite{ranganvamp}. Although there is no theoretical guarantee that VAMP will always converge, strong empirical evidence suggests that VAMP is more resilient to badly conditioned sensing matrices given that they are right-orthogonally invariant \cite{ranganvamp}. 
Consider the joint probability distribution function (pdf) of $\xn$ and $\zn$, $p_{\xn,\zn}(\xs,\zs)$
\begin{equation}
 p_{\xn,\zn}(\xs,\zs)=p_{\xn}(\xs)\CG\N(\zs;\A\xs,\gamma?w^{-1}\I_M).
\end{equation}
Here $p_{\xn}(\xs)$ is some prior distribution on the vector $\xs$ whose elements are assumed to be i.i.d. with a common prior distribution, $p_{\xsn}(x)$, i.e.:
\begin{equation}
p_{\xn}(\xs)=\ds\prod_{i=1}^N p_{\xsn}(x_i).
\end{equation}
Max-sum VAMP can solve the following optimization problem:
\begin{equation}
\label{eqn:reg-slr}
\widehat{\xs}=\ds\argmin_\xs \norm{\zs-\A\xs}^2,
\end{equation}
by finding the \textit{maximum a posteriori} (MAP) estimate of $\xs$ as follows:
\begin{equation}
\widehat{\xs}=\argmax_\xs p_{\bm{\mathsf{x}}|\bm{\mathsf{z}}}(\xs|\zs).
\end{equation}
The algorithm consists of the following two modules.
\subsubsection{Linear MAP/MMSE Estimator}
At iteration $t$, the linear MAP estimator receives extrinsic information (message) from the separable (i.e., element-wise)  MAP denoiser of $\xs$ in the form of a mean vector, $\rs_{t-1}$, and a common scalar precision, $\gamma_{t-1}$. Then, under the Gaussian prior, $\CG\N(\xs;\rs_{t-1},\gamma_{t-1}^{-1}\I_N)$, it computes the linear MAP estimate, $\bar{\xs}_t$, along with the associated posterior precision, $\bar{\gamma}_t$, from the linear observations, $\zs=\A\xs+\ws$ on $\xs$. Because we are dealing with Gaussian densities, the linear MAP estimate is equal to the linear MMSE (LMMSE) and given as follows:
\begin{align}
\label{eqn:lmmse-est}
\bar{\xs}_t&~=~\left(\gamma?w\A^\hr\A+\gamma_{t-1}\I_N\right)^{-1}\left(\gamma?w\A^\hr\zs+\gamma_{t-1}\rs_{t-1}\right), \\
\label{eqn:lmmse-var}
\bar{\gamma}_t&~=~N\tn{Tr}\left(\left[\gamma?w\A^\hr\A+\gamma_{t-1}\I_N\right]^{-1}\right)^{-1}.
\end{align}
The extrinsic information on $\xs$ is updated as
$\CG\N(\xs;\bar{\xs}_t,\bar{\gamma}_t^{-1}\I_N)/\CG\N(\xs;\rs_{t-1},\gamma_{t-1}^{-1}\I_N),$
and then sent back in the form of a mean vector, $\widetilde{\rs}_t=\left(\bar{\xs}_t\bar{\gamma}_t-\rs_{t-1}\gamma_{t-1}\right)/\left(\bar{\gamma}_t-\gamma_{t-1}\right)$, and a scalar precision, $\widetilde{\gamma}_t=\bar{\gamma}_t-\gamma_{t-1}$, to the separable MAP denoiser of $\xs$.
The SVD (singular value decomposition) form of VAMP directly computes extrinsic mean vector $\widetilde{\rs}_t$ and scalar precision $\widetilde{\gamma}_t$, and can be readily obtained by substituting $\A=\U\tn{Diag}(\bm{\omega})\V^\hr$ in \eqref{eqn:lmmse-est} and \eqref{eqn:lmmse-var}.
\subsubsection{Separable MAP Denoiser of \texorpdfstring{$\xs$}{Lg}}
This module computes the MAP estimate, $\widehat{\xs}_t$, of $\xs$ from the joint distribution $p_{\xn}(\xs)\CG\N(\xs;\widetilde{\rs}_t,\widetilde{\gamma}^{-1}_t\I_N)$. Because $\xs$ is i.i.d., the MAP estimate can be computed through a component-wise denoising function as follows:
\begin{equation}
\widehat{x}_{i,t} = g_{1,i}(\widetilde{r}_{i,t},\widetilde{\gamma}_t)\triangleq \argmax_{x_i}\left[-\widetilde{\gamma}_t\abs*{x_i-\widetilde{r}_{i,t}}^2+\ln p_{\xsn}(x_i)\right],
\end{equation}
or equivalently,
\begin{equation}
\label{eqn:denoiser}
g_{1,i}(\widetilde{r}_{i,t},\widetilde{\gamma}_t)= \ds\argmin_{x_i}\left[\widetilde{\gamma}\abs*{x_i-\widetilde{r}_{i,t}}^2-\ln p_{\xsn}(x_i)\right].
\end{equation}
\textcolor{dg}{
The derivative of the scalar MAP denoiser w.r.t. $\widetilde{r}_{i,t}$ is given by \cite{ranganvamp}:
\begin{equation}
\label{eqn:denoiser-der}
g_{1,i}'(\widetilde{r}_{i,t},\widetilde{\gamma}_t)~\triangleq~\frac{\partial g_{1,i}(\widetilde{r}_{i,t},\widetilde{\gamma}_t)}{\partial\widetilde{r}_{i,t}}
~=~\frac{1}{2}\left(\frac{\partial g_{1,i}\left(\widetilde{r}_{i,t},\widetilde{\gamma}_t\right)}{\partial\Re\left\lbrace\widetilde{r}_{i,t}\right\rbrace}-\js\frac{\partial g_{1,i}\left(\widetilde{r}_{i,t},\widetilde{\gamma}_t\right)}{\partial\Im\left\lbrace\widetilde{r}_{i,t}\right\rbrace}\right)
~=~\widetilde{\gamma}_t\widehat{\gamma}_t,
\end{equation}
where $\widehat{\gamma}_t$ is the posterior precision. The vector valued denoiser function and its derivaive are defined as follows:
\begin{align}
\label{eqn:vector-den}
\gs_1(\widetilde{\rs}_t,\widetilde{\gamma}_t)&~\triangleq~ \big[g_{1,1}(\widetilde{r}_{1,t},\widetilde{\gamma}_t), \ g_{1,2}(\widetilde{r}_{2,t},\widetilde{\gamma}_t), \cdots, g_{1,N}(\widetilde{r}_{N,t},\widetilde{\gamma}_t)\big]^\tr,\\
\label{eqn:vector-den-der}
\gs'_1(\widetilde{\rs}_t,\widetilde{\gamma}_t)&~\triangleq~ \big[g'_{1,1}(\widetilde{r}_{1,t},\widetilde{\gamma}_t), \ g'_{1,2}(\widetilde{r}_{2,t},\widetilde{\gamma}_t), \cdots, g'_{1,N}(\widetilde{r}_{N,t},\widetilde{\gamma}_t)\big]^\tr.
\end{align}
}\noindent
Similar to the LMMSE module, the MAP denoiser module computes an extrinsic mean vector, $\rs_t=\left(\widehat{\xs}_t\widehat{\gamma}_t-\widetilde{\rs}_t\widetilde{\gamma}_t\right)/\left(\widehat{\gamma}_t-\widetilde{\gamma}_t\right)$, and a scalar precision, $\gamma_t=\widehat{\gamma}_t-\widetilde{\gamma}_t$, and sends them back to the LMMSE module for the next iteration. The process is repeated until convergence.
\textcolor{dg}{
It is worth mentioning that the extrinsic parameters, i.e., the extrinsic mean vector and the scalar precision, calculated by each module act as a Gaussian prior on the succeeding estimate of the adjacent module, thus making VAMP parameter-free.} Another key advantage of VAMP is that it decouples the prior information, $p_{\xn}(\xs)$, and the observations, $p_{\xn|\zn}(\zs|\xs)$, into two separate modules. Moreover, it also enables the denoising function to be separable even if the elements of $\xs$ are correlated in which case the LMMSE module can easily incorporate such correlation information. The steps of the standard max-sum VAMP algorithm are shown in \textbf{Algorithm \ref{algo:std-vamp}}.
\begin{algorithm}
\caption{Max-sum VAMP SVD}
\label{algo:std-vamp}
\mbox{\small Given $\A\in\C^{M \times N}$, $\zs \in \C^M$, a precision tolerance $(\epsilon)$ and a maximum number of iterations $(T?{MAX})$}
\vspace{-20pt}
\begin{algorithmic}[1]
\STATE Initialize $\rs_0$, $\gamma?{0}\geq0$ and $t\leftarrow 1$
\STATE Compute economy-size SVD $\A=\U\tn{Diag}(\bm{\omega})\V^\hr$
\STATE $R?A= \tn{Rank}(\A)= \tn{length}(\bm{\omega})$
\STATE Compute $\widetilde{\zs} = \tn{Diag}(\bm{\omega})^{-1}\U^\hr\zs$
\REPEAT
\STATE \vspace{2pt}// LMMSE SVD Form.
\STATE $\dk_t=\gamma?w\tn{Diag}(\gamma?w\bm{\omega}^2+\gamma_{t-1}\1_{R?{A}})^{-1}\bm{\omega}^2$
\STATE $\widetilde{\rs}_t=\rs_{t-1}+\frac{N}{R?A}\V\tn{Diag}\left(\dk_t/\langle\dk_t\rangle\right)\left(\widetilde{\zs}-\V^\hr\rs_{t-1}\right)$
\STATE $\widetilde{\gamma}_t=\gamma_{t-1}\left\langle\dk_t\right\rangle/\left(\frac{N}{R?A}-\left\langle\dk_t\right\rangle\right)$ \vspace{5pt}

\STATE // MAP Denoiser
\STATE $\widehat{\xs}_t=\gs_1(\widetilde{\rs}_t,\widetilde{\gamma}_t)$
\STATE $\widehat{\gamma}_t=\left\langle \gs_1'(\widetilde{\rs}_t,\widetilde{\gamma}_t)\right\rangle/\widetilde{\gamma}_t$
\STATE $\gamma_t=\widehat{\gamma}_t-\widetilde{\gamma}_t$
\STATE $\rs_t=(\widehat{\gamma}_t\widehat{\xs}_t-\widetilde{\gamma}_t\widetilde{\rs}_t)/\gamma_t$
\STATE $t\leftarrow t+1$
\vspace{2pt}
\UNTIL $\norm{\widehat{\xs}_t-\widehat{\xs}_{t-1}}_2^2\leq \epsilon\norm{\widehat{\xs}_{t-1}}_2^2$ or $t>T?{MAX}$
\RETURN $\widehat{\xs}_t$
\end{algorithmic}
\end{algorithm}
\vspace{-5pt}
\subsection{VAMP for Optimization}
\label{sec:vamp-opt}
In this section, we explain how max-sum VAMP can be applied to constrained optimization problems. Given the knowledge of three matrices $\A \in \C^{M \times N}$, $\B \in \C^{Q \times N}$ and $\z\in\C^{M \times Q}$, the goal is to solve an optimization problem of the form:
\textcolor{dg}{
\begin{subequations}
\label{eqn:matrix-opt}
\begin{align}
\label{eqn:matrix-opta}
\ds\argmin_{\xs~\in~\C^N} \quad & \norm{\A\tn{Diag}(\xs)\B^\tr-\z}?F^2\\
\label{eqn:matrix-optb}
\tn{s.t.} \quad & f_i(x_{i}) =0 \quad i=1, \cdots, N.
\end{align}
\end{subequations}
}
\noindent
In the context of optimization, the observation matrix, $\z$, is considered as the desired output matrix and it is also assumed to be known. Unlike the estimation problem in \eqref{eqn:reg-slr}, we do not have a prior distribution on $\xs$. Yet, the optimization problem in \eqref{eqn:matrix-opt} can be solved by modifying the modules of standard max-sum VAMP. 

\textcolor{dg}{
\subsubsection{Extended LMMSE}
Through {vectorization}, the objective function in \eqref{eqn:matrix-opta} can be written in the same form as the quadratic objective function in \eqref{eqn:reg-slr} in the following way:
\begin{equation}
\tn{vec}(\z)=(\B\otimes\A)\tn{vec}(\tn{Diag}(\xs)).
\end{equation}
We then define a matrix, $\D \in \C^{MQ \times N}$, as follows:
\begin{equation}
\D\triangleq\B\ast\A=\left[\bs_1 \otimes \as_1, \cdots, \bs_K \otimes \as_K\right].
\end{equation}
Then, the objective function in \eqref{eqn:matrix-opta} is equivalently expressed in a standard form that is amenable to VAMP as follows:
\begin{subequations}
\label{eqn:dmatrix-opt}
\begin{align}
\label{eqn:dmatrix-opta}
\ds\argmin_{\xs~\in~\C^N} \quad & \norm{\D\xs-\tn{vec}(\z)}?2^2\\
\label{eqnd:matrix-optb}
\tn{s.t.} \quad & f_i(x_{i}) =0 \quad i=1, \cdots, N.
\end{align}
\end{subequations}
The column-wise Khatri-Rao structure can be exploited to avoid taking SVD of the large matrix $\D$ as explained in the sequel. Let $\A=\U?A\tn{Diag}(\bm{\omega}?A)\V?A^\hr$, $\B=\U?B\tn{Diag}(\bm{\omega}?B)\V?B^\hr$, \mbox{$\D=\U\tn{Diag}(\bm{\omega})\V^\hr$} and $\V?{BA}=\left(\V?B^\hr \ast \V?A^\hr\right)^\hr$.
By defining the normalization vector: 
\begin{equation}
\vs?n=\big[\norm{\vs_{\tn{BA},1}}_2, \ \norm{\vs_{\tn{BA},2}}_2, \cdots, \norm{\vs_{\tn{BA},MQ}}_2\big]^\tr,
\end{equation}
it can be shown that the SVD of the matrix $\D$ is given by:
\begin{equation}
\D=\underbrace{\left(\U?B \otimes \U?A\right)}_{\U}\underbrace{\tn{Diag}\big((\bm{\omega}?B \otimes \bm{\omega}?A)\odot\vs?n\big)}_{\tn{Diag}(\bm{\omega})}\underbrace{\left(\V?B^\hr \ast \V?A^\hr\right)\odot \left(\vs_{\tn{n}}^{-1}\1_N^\tr\right)}_{\V^\hr}.
\end{equation}
}
These steps can be easily incorporated in the \textbf{Algorithm} \ref{algo:std-vamp} accordingly. Similar to the standard max-sum VAMP, at iteration $t$, the LMMSE module receives an extrinsic mean vector, $\rs_{t-1}$, and a scalar precision, $\gamma_{t-1}$, from the separable MAP estimator. 
The SVD form of VAMP allows for exploiting the Kronecker structure inside the algorithm to avoid any large matrix multiplication. The product of a Kronecker-structured matrix and a vector can be computed in an efficient way through reverse vectorization or \textit{unvectorization} by computing the product of three smaller matrices, and then vectorizing the result. Therefore, line $4$ of \textbf{Algorithm} \ref{algo:std-vamp} can be modified as follows:
\textcolor{dg}{
\begin{equation}
\widetilde{\zs}~=~\tn{Diag}(\bm{\omega})^{-1}\U^\hr\tn{vec}(\z)~=~\tn{Diag}(\bm{\omega})^{-1}\tn{vec}\left(\U?A^\hr\z\U?B^*\right).
\end{equation}
}
\begin{figure}
\vspace{-20pt}
\bc
\scalebox{1.2}{
\begin{picture}(250,110)
\put(0,0){\includegraphics[scale=0.45]{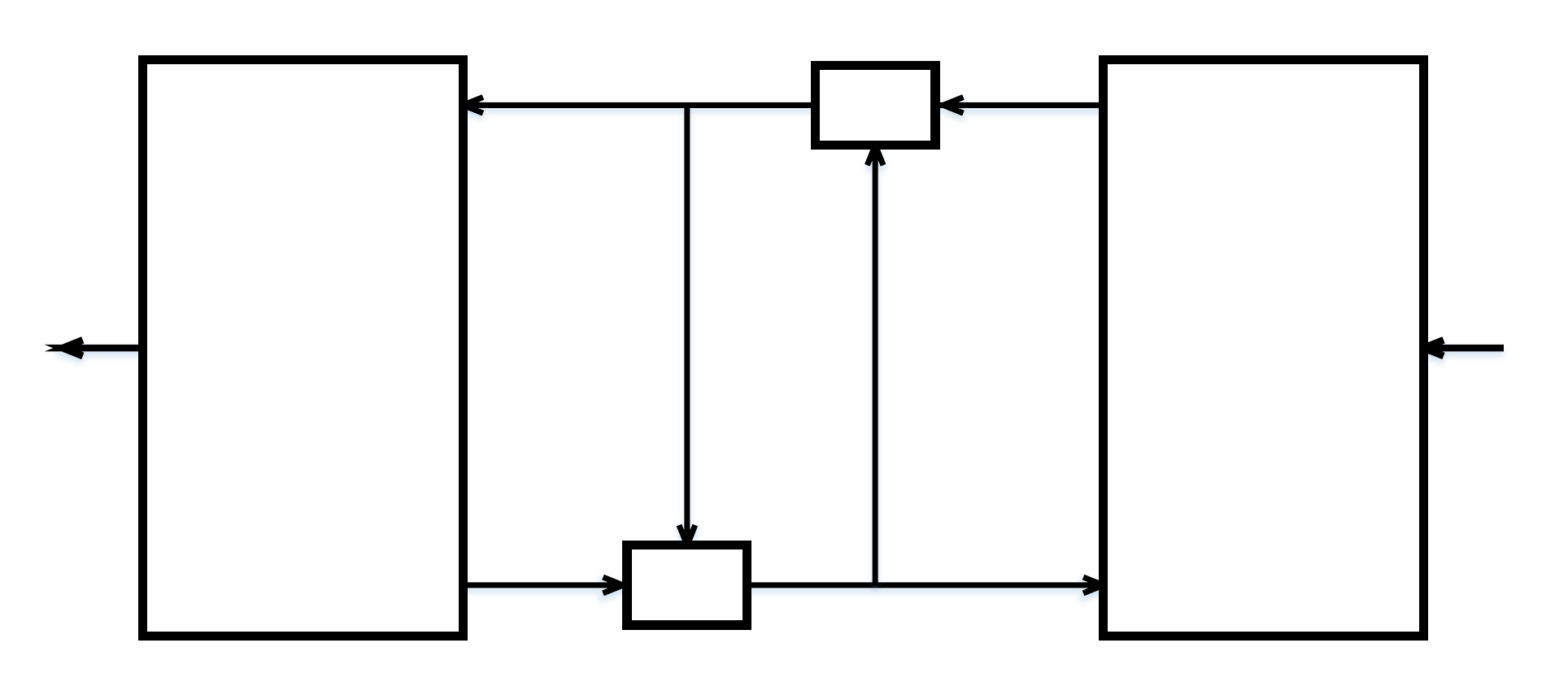}}
\put(34,62){{\footnotesize Separable}}
\put(40,50){{\footnotesize MAP}}
\put(34,40){{\footnotesize Projector}}
\put(188,58){{\footnotesize LMMSE}}
\put(187,47){{\footnotesize Estimator}}
\put(135,92){{\footnotesize ext}}
\put(105,15){{\footnotesize ext}}
\put(12,58){{\footnotesize $\widehat{\xs}$}}
\put(230,58){{\footnotesize $\z$}}
\put(102,98){{\footnotesize $\widetilde{\rs},\widetilde{\gamma}$}}
\put(164,84){{\footnotesize $\bar{\xs}$}}
\put(164,98){{\footnotesize $\bar{\gamma}$}}
\put(134,8){{\footnotesize $\rs,\gamma$}}
\put(80,6){{\footnotesize $\widehat{\xs}$}}
\put(80,22){{\footnotesize $\widehat{\gamma}$}}
\end{picture}
}
\ec
\caption{Block diagram of batch VAMP for optimization. \textcolor{dg}{The calculation of extrinsic information is performed by the ``ext" blocks.}}
\label{fig:vamp-opt}
\vspace{-20pt}
\end{figure}
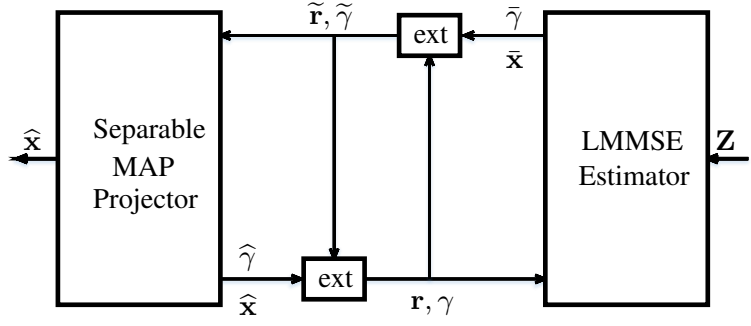
\noindent
The steps for computing the extrinsic mean vector, $\widetilde{\rs}_t$, and the scalar precision, $\widetilde{\gamma}_t$, remain unchanged and they are computed directly without the need for computing the LMMSE estimate, $\bar{\xs}_t$, and the posterior precision, $\bar{\gamma}_t$.
Hence, the only Kronecker product required for the LMMSE is of the two vectors $\bm{\omega}?B$ and $\bm{\omega}?A$.
\begin{algorithm}
\caption{Max-sum VAMP SVD for optimization}
\label{algo:opt-vamp}
\mbox{\small Given $\A\in\C^{M \times N}$, $\B \in \C^{Q \times N}$, $\z \in \C^{M \times Q}$, a precision tolerance $(\epsilon)$ and a maximum number of iterations $(T?{MAX})$}
\vspace{-20pt}
\begin{algorithmic}[1]
\STATE Select initial $\rs_0$, $\gamma_0\geq0$ and $t \leftarrow 1$
\STATE Compute economy-size SVD $\A=\U?A\tn{Diag}(\bm{\omega}?A)\V?A^\hr$
\STATE Compute economy-size SVD $\B=\U?B\tn{Diag}(\bm{\omega}?B)\V?B^\hr$
\STATE Compute $\V?{BA}=\left(\V?B^\hr \ast \V?A^\hr\right)^\hr$
\STATE Compute normalization vector $\vs?n=\big[\norm{\vs_{\tn{BA},1}}_2, \ \norm{\vs_{\tn{BA},2}}_2, \cdots, \norm{\vs_{\tn{BA},MQ}}_2\big]^\tr$
\STATE Compute $\V^\hr=\V?{BA}^\hr\odot \left(\vs_{\tn{n}}^{-1}\1_N^\tr\right)$
\STATE Compute $\bm{\omega} = (\bm{\omega}?B \otimes \bm{\omega}?A)\odot\vs?n$
\STATE Compute $\widetilde{\zs} =\tn{Diag}(\bm{\omega})^{-1}\tn{vec}\left(\U?A^\hr\z\U?B^*\right)$
\STATE $R?{BA}= \tn{Rank}(\B\ast\A)= \tn{length}(\bm{\omega})$
\REPEAT
\STATE \vspace{2pt}// LMMSE SVD Form.
\STATE $\dk_t=\gamma?w\tn{Diag}(\gamma?w\bm{\omega}^2+\gamma_{t-1}\1_{R?{BA}})^{-1}\bm{\omega}^2$
\STATE $\widetilde{\rs}_t=\rs_{t-1}+\frac{N}{R?{BA}}\V\tn{Diag}\left(\dk_t/\langle\dk_t\rangle\right)\left(\widetilde{\zs}-\V^\hr\rs_{t-1}\right)$
\STATE $\widetilde{\gamma}_t=\gamma_{t-1}\left\langle\dk_t\right\rangle/\left(\frac{N}{R?{BA}}-\left\langle\dk_t\right\rangle\right)$ \vspace{5pt}
\STATE // Separable MAP Projector.
\STATE $\widehat{\xs}_t=\gs(\widetilde{\rs}_t,\widetilde{\gamma})$
\STATE $\widehat{\gamma}_t=\widetilde{\gamma}_t^{-1}\left\langle \gs'(\widetilde{\rs}_t,\widetilde{\gamma})\right\rangle$
\STATE $\gamma_t=\widehat{\gamma}_t-\widetilde{\gamma}_t$
\STATE $\rs_t=\gamma_t^{-1}(\widehat{\gamma}_t\widehat{\xs}_t-\widetilde{\gamma}_t\widetilde{\rs}_t)$ 
\STATE $t\leftarrow t+1$
\vspace{2pt}
\UNTIL $\norm{\widehat{\xs}_t-\widehat{\xs}_{t-1}}?2^2\leq \epsilon\norm{\widehat{\xs}_{t-1}}?2^2$ or $t>T?{MAX}$
\RETURN $\widehat{\xs}_t$
\end{algorithmic}
\end{algorithm}
\subsubsection{Scalar MAP Projector}
Because the constraint on $\xs$ is component-wise, we model the constraint on its entries, $x_{i}$, as a prior with some precision, $\gamma?p$, i.e., $p_{\xsn}(x_{i}) \varpropto \exp \left(-\gamma?p\abs*{f_i(x_{i})}^2\right)$ with $\gamma?p \to \infty$. We then define the scalar denoising function (now called projector function in the context of optimization) as follows:
\begin{equation}
\widehat{x}_{i,t}~=~g_{i}(\widetilde{r}_{i,t},\widetilde{\gamma}_t)~\triangleq~ \ds\argmin_{x_{i}}\left[\widetilde{\gamma}_t\abs*{x_{i}-\widetilde{r}_{i,t}}^2-\ln p_{\xsn}(x_{i})\right],
\end{equation}
or equivalently:
\begin{equation}
\label{eqn:gen-projector}
g_{i}(\widetilde{r}_{i,t},\widetilde{\gamma}_t)= \argmin_{x_{i}}\left[\widetilde{\gamma}_t\abs*{x_{i}-\widetilde{r}_{i,t}}^2+\gamma?p\abs*{f_i(x_{i})}^2\right].
\end{equation}
The parameter $\gamma?p$ in \eqref{eqn:gen-projector} accounts for the weight given to the prior on $x_i$ inside the scalar MAP optimization. Therefore, taking $\gamma_p \to \infty$ enforces the constraint. Taking the derivative of the scalar projector function w.r.t. $\widetilde{r}_{i,t}$ as defined in equation \eqref{eqn:denoiser-der} yields:
\begin{equation}
g_{i}'(\widetilde{r}_{i,t},\widetilde{\gamma}_t)=\widetilde{\gamma}_t\widehat{\gamma}_t,
\end{equation}
where $\widehat{\gamma}_t$ is the posterior precision. \textcolor{dg}{The vector valued projector function, $\gs(\widetilde{\rs}_t,\widetilde{\gamma})$, and its derivative, $\gs'(\widetilde{\rs}_t,\widetilde{\gamma})$, are defined in the same way as \eqref{eqn:vector-den} and \eqref{eqn:vector-den-der} respectively.} Similar to the denoiser module, extrinsic information from the projector module is calculated in the form of the mean vector,
$\rs_t=\left(\widehat{\xs}_t\widehat{\gamma}_t-\widetilde{\rs}_t\widetilde{\gamma}_t\right)/(\widehat{\gamma}_t-\widetilde{\gamma}_t)$, and scalar precision, $\gamma_t=\widehat{\gamma}_t-\widetilde{\gamma}_t$, which are then fed to the LMMSE module. In an analogous way to sum-product VAMP, the max-sum VAMP (for optimization) decouples the constraint from the objective function and also enables the projector function to be separable. While the LMMSE module optimizes the objective function with no constraints, the latter are enforced by the projector function. This modular property makes VAMP a robust algorithm for solving optimization problems in the presence of linear mixing and under various component-wise constraints. The block diagram and the algorithmic steps for the optimization-oriented VAMP are presented in Fig. \ref{fig:vamp-opt} and \textbf{Algorithm \ref{algo:opt-vamp}} respectively.

\section{VAMP-Based Solution for the Joint Beamforming Problem}
\label{sec:joint-phase-precod}
In this section, we apply the optimization-oriented VAMP algorithm, described in Section \ref{sec:vamp-opt}, to simultaneously optimize the matrix of phase shifters, ${\mathbf\Upsilon}$, as well as the optimal precoding matrix $\f$. We decouple the joint optimization problem into two sub-problems through alternate optimization. In one side we optimize ${\mathbf\Upsilon}$ by utilizing the modified max-sum VAMP and, on the other side, we find the optimal transmit precoding $\f$.
\vspace{-10pt}
\subsection{Alternating Optimization}
We use alternating minimization which is the two-block version of the block coordinate descent (BCD) algorithm. It is a simple iterative approach that optimizes one variable at a time\footnote{Note here that a variable can be a scalar, a vector, or a whole matrix.} (while fixing the others) and the process is repeated for every variable. \textcolor{dg}{Although it is hard to analytically establish the optimality of the alternating minimization technique for non-convex optimization problems, it is known that it performs really well for various non-convex optimization problems especially for large system sizes \cite{altmin1,altmin2,altmin3,altmin4}.} More specifically, we divide the optimization problem in \eqref{eqn:mmse-opt} into the following two sub-optimization problems:
\begin{enumerate}
\item \vspace{-10pt}
\begin{subequations}
\label{eqn:ao-phase-mat}
\begin{align}
\label{eqn:ao-phase-mata}
\ds\argmin_{{\mathbf\Upsilon}} \quad & \E_{\ys,\s}\big\{\norm{\ys-\s}?2^2\big\}\\
\label{eqn:ao-phase-matb}
\tn{s.t.} \quad & \upsilon_{ik}=0, \quad i \neq k, \\
\label{eqn:ao-phase-matc}
& |\upsilon_{ii}|=1, \quad i=1,2, \cdots, K.
\end{align}
\end{subequations}
\item \vspace{-20pt}
\begin{subequations}
\label{eqn:ao-precod-mat}
\begin{align}
\label{eqn:ao-precod-mata}
\argmin_{\alpha, \f} \quad & \E_{\ys,\s}\big\{\norm{\ys-\s}?2^2\big\}\\
\label{eqn:ao-precod-matb}
\tn{s.t.} \quad & \E_{\s}\|\f\s\|?2^2 =P.
\end{align}
\end{subequations}
\end{enumerate}
Let us define the error at iteration $t$ as follows:
\begin{equation}
E_t~\triangleq~\norm{\widehat{\alpha}_t\left(\h?{s-u}^\hr\widehat{{\mathbf\Upsilon}}_t\h?{b-s}+\h?{b-u}^\hr\right)\widehat{\f}_t-\I_M}?F^2 + M\widehat{\alpha}_t^2\sigma?w^2.
\end{equation}
The algorithm stops iterating when $|E_t-E_{t-1}|<\epsilon E_{t-1}$, where $\epsilon \in \R_+$ is some precision tolerance. The algorithmic steps for alternating minimization (after evaluating the expectation) are shown in \textbf{Algorithm \ref{algo:bcd}}.
\begin{algorithm}
\caption{Alternating minimization}
\label{algo:bcd}
\mbox{Given $\h?{s-u}$, $\h?{b-u}$, $\h?{b-s}$, a precision tolerance $(\epsilon)$, and a maximum number of iterations $(T?{MAX})$}
\vspace{-20pt}
\begin{algorithmic}[1]
\STATE Initialize $\widehat{{\mathbf\Upsilon}}?0$ and $t \leftarrow 1$.
\REPEAT \vspace{-20pt}
\STATE \begin{align*}
[\widehat{\alpha}_t,\widehat{\f}_t]= \ds\argmin_{\alpha, \f} & \quad \norm{\alpha\left(\h?{s-u}^\hr\widehat{{\mathbf\Upsilon}}_{t-1}\h?{b-s}+\h?{b-u}^\hr\right)\f-\I_M}?F^2 + M\alpha^2\sigma?w^2\\
\quad \tn{s.t.} & \quad \|\f\|?F^2 = P 
\end{align*} \vspace{-40pt}
\STATE \begin{align*}
\widehat{{\mathbf\Upsilon}}_t=\argmin_{{\mathbf\Upsilon}} \quad & \norm{\widehat{\alpha}_t\left(\h?{s-u}^\hr{\mathbf\Upsilon}\h?{b-s}+\h?{b-u}^\hr\right)\widehat{\f}_t-\I_M}?F^2\\
\tn{s.t.} \quad & \upsilon_{ik}=0 \quad  i \neq k,\\
& |\upsilon_{ii}|=1 \quad  i=1,2, \cdots, K
\end{align*}\vspace{-30pt}
\STATE $t \leftarrow t+1$
\UNTIL $|E_t-E_{t-1}|<\epsilon E_{t-1}$ or $t>T?{MAX}$
\RETURN $\widehat{{\mathbf\Upsilon}}_t, \ \widehat{\f}_t, \ \widehat{\alpha}_t $.
\end{algorithmic}
\end{algorithm}

\subsection{Optimization of the Phase Shifters Matrix}

\label{sec:sub-opt-phase}
Here, we specialize optimization-oriented VAMP algorithm introduced in Section \ref{sec:vamp-opt} in order to optimize the phase matrix, ${\mathbf\Upsilon}$. Let us restate the associated optimization after explicitly finding the expectation  in \eqref{eqn:ao-phase-mat} as follows:
\textcolor{dg}{
\begin{subequations}
\label{eqn:opt-phase-mat}
\begin{align}
\label{eqn:opt-phase-mata}
\ds\argmin_{{\bm\upsilon}} \quad & \norm{\alpha\h?{s-u}^\hr\tn{Diag}({\bm\upsilon})\h?{b-s}\f-(\I_M-\alpha\h?{b-u}^\hr\f)}?F^2\\
\label{eqn:opt-phase-matb}
\tn{s.t.} \quad & |\upsilon_{i}|=1 \quad  i=1,2, \cdots, K.
\end{align}
\end{subequations}
}
The solution is obtained by setting $\A=\alpha\h?{s-u}^\hr$, $\B=(\h?{b-s}\f)^\tr$ and $\z=\I_M-\alpha\h?{b-u}\f$ in \textbf{Algorithm \ref{algo:opt-vamp}} and then choosing a suitable projector function to satisfy the constraints on the reflection coefficients. The unconstrained minimization of the objective function in \eqref{eqn:opt-phase-mata} is performed by the LMMSE module. We define the projector function that enforces the constraint on the reflection coefficients as:
\textcolor{dg}{
\begin{equation}
\label{eqn:uni-mod-projector-opt}
g_{2,i}\left(\widetilde{r}_{i},\widetilde{\gamma}\right)~\triangleq~
  \ds\argmin_{\upsilon_{i}}\left[\widetilde{\gamma}\abs*{\upsilon_{i}-\widetilde{r}_{i}}^2+\gamma?p\big|\abs*{\upsilon_{i}}-1\big|^2\right].
\end{equation}
}\noindent
Solving the optimization problem in \eqref{eqn:uni-mod-projector-opt} results in the following closed-form expression for the underlying projector function:
\textcolor{dg}{
\begin{equation}
g_{2,i}\left(\widetilde{r}_{i},\widetilde{\gamma}\right)~=~\frac{\widetilde{\gamma}}{\widetilde{\gamma}+\gamma?p}\widetilde{r}_{i}~+~\frac{\gamma?p}{\widetilde{\gamma}+\gamma?p}\widetilde{r}_{i}\abs*{\widetilde{r}_{i}}^{-1}.
\end{equation}
}\noindent
As $\gamma?p \to \infty$, we have, $\frac{\widetilde{\gamma}}{\widetilde{\gamma}+\gamma?p} \to 0$ and $\frac{\gamma?p}{\widetilde{\gamma}+\gamma?p} \to 1$. Therefore, the projector function simplifies to:
\textcolor{dg}{
\begin{equation}
\label{eqn:uni-mod-projector}
g_{2,i}\left(\widetilde{r}_{i}\right)~=~
  \widetilde{r}_{i}\abs*{\widetilde{r}_{i}}^{-1}.
\end{equation}
}\noindent
The derivative of the projector function \eqref{eqn:uni-mod-projector} w.r.t. $\widetilde{r}_{i}$ is obtained according to equation \eqref{eqn:denoiser-der} as follows:
\textcolor{dg}{
\begin{equation}
g'_{2,i}\left(\widetilde{r}_{i}\right)=
  \frac{1}{2}\abs*{\widetilde{r}_{i}}^{-1}.
\end{equation}
}\noindent
\textcolor{dg}{Finally, the projector function, $\gs_2(\widetilde{\rs}_t)$, and its derivative $\gs_2'(\widetilde{\rs}_t)$ are obtained by following \eqref{eqn:vector-den} and \eqref{eqn:vector-den-der} respectively.}
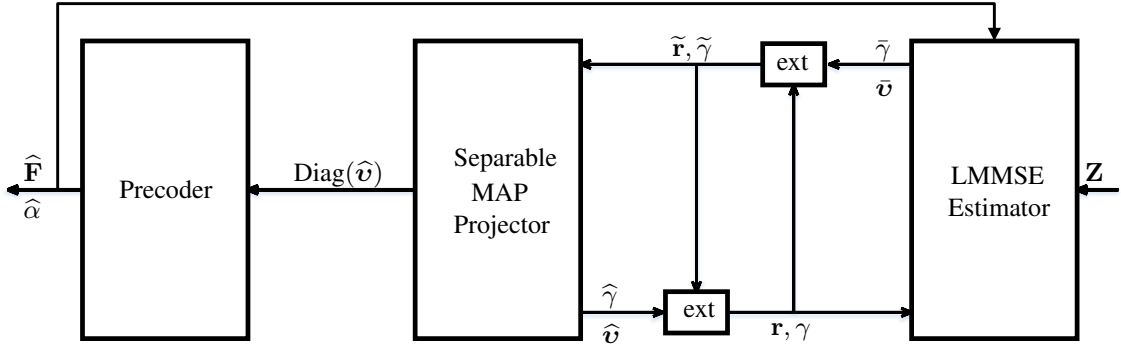
\begin{figure}
\vspace{-20pt}
\bc
\scalebox{1.1}{
\begin{picture}(500,150)(-7.5,0)
\put(0,0){\includegraphics[scale=0.5]{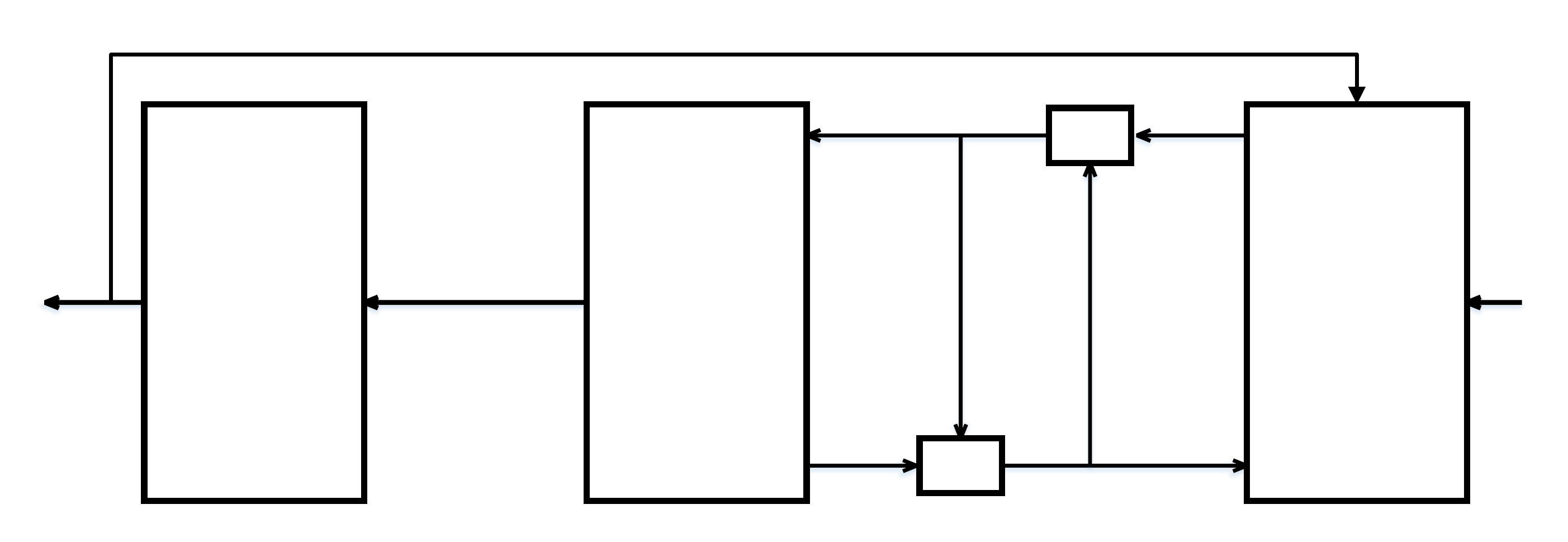}}
\put(165,70){{\footnotesize Separable}}
\put(171,58){{\footnotesize MAP}}
\put(165,47){{\footnotesize Projector}}
\put(336,64){{\footnotesize LMMSE}}
\put(335,53){{\footnotesize Estimator}}
\put(276,103){{\footnotesize ext}}
\put(244,18){{\footnotesize ext}}
\put(110,65){{\footnotesize $\tn{Diag}(\widehat{{\bm\upsilon}})$}}
\put(382,65){{\footnotesize $\z$}}
\put(240,108){{\footnotesize $\widetilde{\rs},\widetilde{\gamma}$}}
\put(310,108){{\footnotesize $\bar{\gamma}$}}
\put(310,94){{\footnotesize $\bar{{\bm\upsilon}}$}}
\put(274,12){{\footnotesize $\rs,\gamma$}}
\put(216,9){{\footnotesize $\widehat{{\bm\upsilon}}$}}
\put(216,23){{\footnotesize $\widehat{\gamma}$}}
\put(50,59){{\footnotesize Precoder}}
\put(17,65){{\footnotesize $\widehat{\f}$}}
\put(17,52){{\footnotesize $\widehat{\alpha}$}}
\end{picture}
}
\ec
\caption{Block diagram of the proposed algorithm.}
\label{fig:overall}
\vspace{-20pt}
\end{figure}
\vspace{-20pt}
\subsection{Optimal Precoding}

The sub-optimization problem in \eqref{eqn:ao-precod-mat} is a constrained MMSE transmit precoding optimization for traditional MIMO systems. It can be solved by jointly optimizing $\f$ and $\alpha$ using Lagrange optimization. After finding the expectation, we construct the Lagrangian function associated to the problem in \eqref{eqn:ao-precod-mat} as follows:
\begin{equation}
\LA(\f,\alpha,\lambda)=\norm{\alpha\h^\hr\f-\I_M}?F^2 + M\alpha^2\sigma?w^2 + \lambda(\tn{Tr}(\f\f^\hr)-P),
\end{equation}
with $\lambda \in \R $ being the Lagrange multiplier.
The closed-form solutions for optimal $\alpha$ and $\f$ are given below and we refer the reader to \cite{nossek2005} for more details: 
\begin{equation}
\label{eqn:opt-alpha}
\alpha\<{opt}~=~g_3\left(\h\right)~\triangleq~\sqrt{\frac{1}{P}}\sqrt{\tn{Tr}\left(\left[\h\h^\hr+\frac{M\sigma?w^2\I_N}{P}\right]^{-2}\h\h^\hr\right)}.
\end{equation}
\begin{equation}
\label{eqn:opt-precod}
\f\<{opt}~=~g_4\left(\h\right)\triangleq\frac{\sqrt{P}\left[\h\h^\hr+\frac{M\sigma?w^2\I_N}{P}\right]^{-1}\h}{\sqrt{\tn{Tr}\left(\left[\h\h^\hr+\frac{M\sigma?w^2\I_N}{P}\right]^{-2}\h\h^\hr\right)}}~=~\alpha^{\tn{opt}^{-1}}\left[\h\h^\hr+\frac{M\sigma?w^2\I_N}{P}\right]^{-1}\h.
\end{equation}
\textcolor{dg}{
Note that, the scalar, $\alpha$, merely represents a scaling factor at the receiver that is used to scale the incident signal as so to obtain the transmitted constellation symbols and this a common practice in MMSE precoding optimization \cite{nossek2005,jeddaprecoding}. Choosing a common $\alpha$ for all users results in better tractability and makes it possible to derive a closed-form solution for the optimal $\alpha$. By inspecting the closed-form solution of the precoding matrix, we observe that it is scaled by $\alpha^{\tn{opt}^{-1}}$. This allows the transmitter to optimally scale all the transmit symbols based on the available transmit power whereas the receiver upscales the received signal plus noise to get back the original transmitted symbols while keeping the SNR unaffected. It is also worth mentioning that the optimal scaling factor, $\alpha\<{opt}$, is only utilized in optimizing the precoding matrix since the receivers can blindly estimate this scalar based on the received symbol sequence \cite{nossek2005,jeddaprecoding}.
}
Now that we have solved both sub-optimization problems in \eqref{eqn:ao-phase-mat} and \eqref{eqn:ao-precod-mat}, separately, we substitute their solutions into \textbf{Algorithm \ref{algo:bcd}}. The overall block diagram and algorithmic steps are respectively shown in Fig. \ref{fig:overall} and \textbf{Algorithm \ref{algo:overall}}.
\textcolor{dg}{
\begin{remark}
It is possible to include per-user data requirement by formulating the problem under the weighted MMSE (WMMSE) criterion where we scale the MSEs of the users with weights according to each user's data requirement and then minimize the sum MSE. To that end, we define a positive semi-definite real diagonal matrix, $\q$, containing user weights, $\{q_m\}_{m=1}^M$, in its diagonal, i.e., $\q=\tn{Diag}(q_1, \cdots, q_M)$. Then, the optimization problem under the WMMSE criterion is given by:
\begin{subequations}
\begin{align}
\ds\argmin_{\alpha, \f, {\mathbf\Upsilon}} \quad & \E_{\ys,\s}\left\lbrace\norm{\q^{1/2}(\ys-\s)}_2^2\right\rbrace,\\
\tn{subject to} \quad & \E_{\s}\left\lbrace\|\f\s\|_2^2\right\rbrace=P,\\
& \upsilon_{ik}=0, \quad i \neq k,\\
& |\upsilon_{ii}|=1, \quad i=1,2, \cdots, K.
\end{align}
\end{subequations}
In this formulation, WMMSE precoding optimization is performed instead of the ordinary MMSE precoding optimization, wherein the matrices $\A$, $\B$ and $\z$ are adjusted accordingly inside the VAMP part of the \textbf{Algorithm \ref{algo:overall}}.
\end{remark}
}
\begin{algorithm*}
\caption{VAMP-based joint optimization algorithm}
\label{algo:overall}
\mbox{Given $\h?{s-u}$, $\h?{b-u}$, $\h?{b-s}$, a precision tolerance $(\epsilon)$, and a maximum number of iterations $(T?{MAX})$}
\vspace{-20pt}
\begin{algorithmic}[1]
\STATE Initialize $\widehat{{\bm\upsilon}}?0, \ \rs_0$, $\gamma_0\geq0$ and $t\leftarrow 1$
\STATE $\widehat{\h}_0=\left(\h?{s-u}^\hr\tn{Diag}(\widehat{{\bm\upsilon}}?0)\h?{b-s} +\h?{b-u}^\hr\right)^\hr$
\STATE $\widehat{\alpha}?0=g_3\left(\widehat{\h}_0\right)$
\STATE $\widehat{\f}?0=g_4\left(\widehat{\h}_0\right)$ \vspace{2pt}
\REPEAT
\STATE // LMMSE SVD Form.
\STATE Set $\A= \widehat{\alpha}_{t-1}\h?{s-u}^\hr$, $\B=\left(\h?{b-s}\widehat{\f}_{t-1}\right)^\tr$ and $\z=\I_M-\widehat{\alpha}_{t-1}\h?{b-u}^\hr\widehat{\f}_{t-1}$.
\STATE Compute economy-size SVD $\A=\U?A\tn{Diag}(\bm{\omega}?A)\V?A^\hr$
\STATE Compute economy-size SVD $\B=\U?B\tn{Diag}(\bm{\omega}?B)\V?B^\hr$
\STATE Compute $\V?{BA}=\left(\V?B^\hr \ast \V?A^\hr\right)^\hr$
\STATE Compute normalization vector $\vs?n=\big[\norm{\vs_{\tn{BA},1}}_2, \ \norm{\vs_{\tn{BA},2}}_2, \cdots, \norm{\vs_{\tn{BA},M^2}}_2\big]^\tr$
\STATE Compute $\V^\hr=\V?{BA}^\hr\odot \left(\vs_{\tn{n}}^{-1}\1_K^\tr\right)$
\STATE Compute $\bm{\omega} = (\bm{\omega}?B \otimes \bm{\omega}?A)\odot\vs?n$
\STATE Compute $\widetilde{\zs} ={\bm\omega}^{-1}\odot\tn{vec}\left(\U?A^\hr\z\U?B^*\right)$
\STATE $R?{BA}= \tn{Rank}(\B\ast\A)= \tn{length}(\bm{\omega})$
\STATE $\dk_t=\gamma?w(\gamma?w\bm{\omega}^2+\gamma_{t-1}{\mathbf 1_{R?{BA}}})^{-1}\odot\bm{\omega}^2$
\STATE $\widetilde{\rs}_t=\rs_{t-1}+\frac{K}{R?{BA}}\V\left(\frac{\dk_t}{\langle\dk_t\rangle}\odot\left(\widetilde{\zs}-\V^\hr\rs_{t-1}\right)\right)$
\STATE $\widetilde{\gamma}_t=\gamma_{t-1}\left\langle\dk_t\right\rangle/\left(\frac{K}{R?{BA}}-\left\langle\dk_t\right\rangle\right)$ \vspace{4pt}
\STATE // Separable MAP Projector
\STATE $\widehat{{\bm\upsilon}}_t=\gs_2\left(\widetilde{\rs}_t\right)$
\STATE $\widehat{\gamma}_t=\widetilde{\gamma}_t^{-1}\left\langle \gs'_2\left(\widetilde{\rs}_t\right)\right\rangle$.
\STATE $\gamma_t=\widehat{\gamma}_t-\widetilde{\gamma}_t$
\STATE $\rs_t=\gamma_t^{-1}\left(\widehat{\gamma}_t\widehat{{\bm\upsilon}}_t-\widetilde{\gamma}_t\widetilde{\rs}_t\right)$ \vspace{4pt}
\STATE //Find $\alpha$ and $\f$ through their closed-form solutions.
\STATE $\widehat{\h}_t=\left(\h?{s-u}^\hr\tn{Diag}(\widehat{{\bm\upsilon}}_t)\h?{b-s} +\h?{b-u}^\hr\right)^\hr$
\STATE $\widehat{\alpha}_t=g_3\left(\widehat{\h}_t\right)$
\STATE $\widehat{\f}_t=g_4\left(\widehat{\h}_t\right)$
\STATE $t \leftarrow t+1$
\UNTIL $|E_t-E_{t-1}|<\epsilon E_{t-1}$ or $t>T?{MAX}$
\RETURN $\widehat{{\bm\upsilon}}_t, \ \widehat{\f}_t, \ \widehat{\alpha}_t$.
\end{algorithmic}
\end{algorithm*}
\vspace{-10pt}
\section{Joint Beamforming Under Reactive Loading at the IRS}
\textcolor{dg}{
\label{sec:reactive-loading}
We consider a reflective element that is combined with a tunable reactive load \footnote{This can be implemented for instance by an antenna array composed of omni-directional dipole elements loaded with the reactive elements in the absence
of a ground plane to allow for bidirectional beamforming and not just  hemispherical coverage.} instead of an ideal phase shifter, i.e., \footnote{The value $1$ is the normalized resistive part of the element impedance whereas $\chi_{i}$ is the normalized reactive part of the antenna plus reactive termination. Accordingly $\upsilon_{i}$ represents the induced current flowing across the antenna. We assume the antenna elements to be uncoupled which holds approximately for half-wavelength element spacing.} $\upsilon_{i}=-(1+\js\chi_{i})^{-1}$,  where $\chi_{i} \in \R$ is a scalar reactance value that has to be optimized for each reflection coefficient. Under the unimodular constraint, the idealistic IRS has a full field of view (FOV) and the reflection coefficients correspond to ideal phase-shifters and are of the form $\upsilon_i=e^{\js\theta_i}$, where $\theta_i \in [0, 2\pi]$\footnote{In practice, this assumption is difficult from a practical standpoint. With the reactive-loading constraint, the assumption of an IRS with full FOV becomes more acceptable.}, whereas under the practical constraint we have a restriction on the possible values of the IRS phase shifts i.e., $\angle -(1+\js\chi)^{-1} \in \big[\tn{-}\frac{\pi}{2}, \frac{\pi}{2}\big]$. Moreover, the magnitude of each phase shift under this constraint is always less than $1$ for any $\chi\neq 0$. Practically, this introduces the phase-dependent amplitude attenuation in the incident wave. We rewrite the objective function under the new constraint on phases as follows: 
\begin{subequations}
\begin{align}
\ds\argmin_{{\bm\upsilon}} \quad & \norm{\alpha\h?{s-u}^\hr\tn{Diag}({\bm\upsilon})\h?{b-s}\f-(\I_M-\alpha\h?{b-u}^\hr\f)}?F^2\\
\tn{s.t.} \quad & \upsilon_{i} = \frac{-1}{1+\js\chi_{i}}, \quad i=1,2, \cdots, K.
\end{align}
\end{subequations}}\noindent
To find the sub-optimal phase matrix under the new constraint, we change the projector function accordingly as follows:
\textcolor{dg}{
\begin{equation}
\label{eqn:reactive-projector}
g_{5,i}\left(\widetilde{r}_{i},\widetilde{\gamma}\right)~\triangleq~
\ds\argmin_{\upsilon_{i}}\left[\widetilde{\gamma}\abs*{\upsilon_{i}-\widetilde{r}_{i}}^2+\gamma?p\abs*{\upsilon_{i}+\frac{1}{1+\js\chi_{i}\<{opt}}}^2\right],
\end{equation}
}
where
\begin{equation}
\label{eqn:reactance-opt}
\chi_{i}\<{opt}~=~g_6\left(\widetilde{r}_{i}\right)~\triangleq~\ds\argmin_{\chi_{i}} \abs*{\widetilde{r}_{i}+\frac{1}{1+\js\chi_{i}}}^2.
\end{equation}
The optimization problem in \eqref{eqn:reactive-projector} is a bi-level one \cite{sinbilevel}. The solution to \eqref{eqn:reactance-opt} is substituted in \eqref{eqn:reactive-projector} which is then solved as ordinary MAP optimization. We show in \textbf{Appendix \ref{apd:A}} that the solution to \eqref{eqn:reactance-opt} is given by: 
\begin{equation}
\label{eqn:reactance-opt-sol}
g_6\left(\widetilde{r}_{i}\right)~=~
\frac{1}{2\Im\left\lbrace\widetilde{r}_{i}\right\rbrace}\left(1+2\Re\left\lbrace\widetilde{r}_{i}\right\rbrace+\sqrt{\left(1+2\Re\left\lbrace\widetilde{r}_{i}\right\rbrace\right)^2+4\Im\left\lbrace\widetilde{r}_{i}\right\rbrace^2}\right).
\end{equation}
Substituting \eqref{eqn:reactance-opt-sol} back into \eqref{eqn:reactive-projector} and solving the minimization leads to the following result:
\textcolor{dg}{
\begin{equation}
g_{5,i}(\widetilde{r}_{i},\widetilde{\gamma})=\frac{\widetilde{\gamma}}{\widetilde{\gamma}+\gamma?p}\widetilde{r}_{i}-\frac{\gamma?p}{\widetilde{\gamma}+\gamma?p}\left(1+\js g_6\left(\widetilde{r}_{i}\right)\right)^{-1},
\end{equation}
}\noindent
where $\gamma?p \to \infty. $ Thus, the projector function can be expressed as:
\textcolor{dg}{
\begin{equation}
g_{5,i}\left(\widetilde{r}_{i}\right)=
  -\left(1+\js g_6\left(\widetilde{r}_{i}\right)\right)^{-1},
\end{equation}
}\noindent
whose derivative is obtained as defined in equation \eqref{eqn:denoiser-der} as follows:
\textcolor{dg}{
\begin{equation}
\label{eqn:reactive-projector-der}
g'_{5,i}\left(\widetilde{r}_{i}\right)=
  \js g'_6\left(\widetilde{r}_{i}\right)\left(1+\js g_6\left(\widetilde{r}_{i}\right)\right)^{-2},
\end{equation}
}\noindent
where 
\begin{equation}
\label{eqn:reactive-func-der}
 g'_6\left(\widetilde{r}_{i}\right)= \frac{1}{2}\left(\frac{\partial g_6\left(\widetilde{r}_{i}\right)}{\partial\Re\left\lbrace\widetilde{r}_{i}\right\rbrace}-\js\frac{\partial g_6\left(\widetilde{r}_{i}\right)}{\partial\Im\left\lbrace\widetilde{r}_{i}\right\rbrace}\right).
\end{equation}
The partial derivatives involved in \eqref{eqn:reactive-func-der} are given by:
\begin{multline}
\frac{\partial g_6\left(\widetilde{r}_{i}\right)}{\partial\Re\left\lbrace\widetilde{r}_{i}\right\rbrace}=\Im\left\lbrace\widetilde{r}_{i}\right\rbrace^{-1}
+\left(1+2\Re\left\lbrace\widetilde{r}_{i}\right\rbrace\right)\left(\Im\left\lbrace\widetilde{r}_{i}\right\rbrace\sqrt{\left(1+2\Re\left\lbrace\widetilde{r}_{i}\right\rbrace\right)^2+4\Im\left\lbrace\widetilde{r}_{i}\right\rbrace^2}\right)^{-1},
\end{multline}
and
\begin{multline}
\frac{\partial g_6\left(\widetilde{r}_{i}\right)}{\partial\Im\left\lbrace\widetilde{r}_{i}\right\rbrace}=-\left(1+2\Re\left\lbrace\widetilde{r}_{i}\right\rbrace\right)\left(2\Im\left\lbrace\widetilde{r}_{i}\right\rbrace^2\right)^{-1} \\
-\left(1+2\Re\left\lbrace\widetilde{r}_{i}\right\rbrace\right)^2\left(2\Im\left\lbrace\widetilde{r}_{i}\right\rbrace^2\sqrt{\left(1+2\Re\left\lbrace\widetilde{r}_{i}\right\rbrace\right)^2+4\Im\left\lbrace\widetilde{r}_{i}\right\rbrace^2}\right)^{-1}.
\end{multline}
Since the derivative is required to be a real scalar, we take the absolute value of the complex derivative and, therefore, we modify the derivative of the projector function \eqref{eqn:reactive-projector-der} as follows:
\textcolor{dg}{
\begin{equation}
g'_{5,i}\left(\widetilde{r}_{i}\right)=
  \abs*{\js g'_6\left(\widetilde{r}_{i}\right)\left(1+\js g_6\left(\widetilde{r}_{i}\right)\right)^{-2}}.
\end{equation}
}\noindent
\textcolor{dg}{Lastly, we obtain the vector valued projector function, $\gs_5(\widetilde{\rs}_t)$, and its derivative $\gs_5'(\widetilde{\rs}_t)$ according to \eqref{eqn:vector-den}  and \eqref{eqn:vector-den-der} respectively and replace $\gs_2(\widetilde{\rs}_t)$ and $\gs_2'(\widetilde{\rs}_t)$ in lines $19$ and $20$ of \textbf{Algorithm \ref{algo:overall}}.}

\section{Numerical Results: Performance and Complexity Analysis}

\label{sec:numeric-res}

\subsection{Simulation Model and Parameters}

We present exhaustive Monte-Carlo simulation results to assess the performance of the proposed algorithm. We assume that the IRS is located at a fixed distance of $500$ m from the BS and the users are spread uniformly at a radial distance of $10$ m to $50$ m from the IRS. \textcolor{dg}{A path-based propagation channel model, also known as parametric channel model \cite{heathmimo}, is used. Such a model are more appropriate for systems with large antenna arrays. One key parameter of such a channel model is the number of multi-path components of the BS-IRS channel which governs the effect of channel correlation.}
\vspace{-5pt}
\begin{table}[!ht]
\centering
\caption{Simulation parameters, their notations and values.}
\label{table:sim-parameter}
\textcolor{dg}{
\begin{tabular}{|p{0.32\textwidth}|P{0.15\textwidth}|p{0.25\textwidth}|P{0.15\textwidth}|}
\hline 
\vspace{-10pt} \center{\textbf{Parameter}}  &  \vspace{-5pt}\textbf{Notation, Value}  & \vspace{-10pt} \center{\textbf{Parameter}}  &  \vspace{-5pt}\textbf{Notation, Value}\\
 \hline
\hline 
 Number of channel paths IRS-BS link & $Q?{IRS}=10$ & 
 IRS-BS distance & $d?{IRS}=500$ m\\
\hline
 Number of channel paths BS-user link & $Q?{b-u}=2$ &
 User-BS distance & $d=500$ m\\
\hline 
 Number of channel paths IRS-user link & $Q?{s-u}=2$&
 User-IRS distance & $d' \in [10,50]$ m \\
\hline 
 Path-loss exponent IRS-BS, IRS-user link & $\eta=2.5$&
 Noise variance & $\sigma?w^2=-100$ dBm \\
\hline 
 Path-loss exponent BS-user link & $\eta=3.7$&
 Channel path gain & $c_q\sim\CG\N(0,1)$ \\
\hline 
 Reference distance & $d?0=1$ m &
 Path-loss at reference distance & $C?0=-30$ dB \\
\hline
\end{tabular}
}
\end{table}
The channel between the IRS and the BS is generated according to:
\vspace{-10pt}
\begin{equation}
\label{eqn:chanl-b-s}
\h?{b-s}~=~\sqrt{L(d?{IRS})}\ds\sum_{q=1}^{Q?{IRS}}\cn_q\as?{IRS}(\varphi_q,\psi_q)\as?{BS}(\phi_q)^\tr.
\end{equation}
Here, $Q?{IRS}$ and $L(d?{IRS})$ denote the number of channel paths and the distance-dependent path-loss factor, respectively. The vectors $\as?{BS}(\phi)$ and $\as?{IRS}(\varphi,\psi)$ are the array response vectors for the BS and the IRS, respectively. The coefficients $c_q$ in \eqref{eqn:chanl-b-s} denote the path gains which are modeled by a complex normal distribution, i.e., $\cn_q\sim\CG\N(c_q;0,1)$. Assuming that a uniform linear array (ULA) with $N$ antennas is used at the BS, we have
$\as?{BS}(\phi)=[1,e^{2\pi \js\frac{d?b}{\lambda}\cos\phi},\cdots, e^{2\pi \js\frac{d?b}{\lambda}(N-1)\cos\phi}]^\tr$ wherein $\lambda$, $\phi$, and $d?b$ represent the wavelength, the angle of departure (AOD), and the inter-antenna spacing at the BS, respectively. The IRS is equipped with a (square) uniform planar array (UPA) with $K$ antenna elements which are assumed to have a cosine embedded element pattern. By defining the z-axis as the normal vector to the array, the array response vector for the IRS is expressed as follows \cite{hannanant}:
\begin{equation}
\as?{IRS}(\varphi,\psi)~=~
\sqrt{|\cos\varphi|}
\begin{bmatrix}
1\\
e^{2\pi \js\frac{d?s}{\lambda}\sin\varphi\sin\psi}\\
\vdots\\
e^{2\pi \js\frac{d?s}{\lambda}(\sqrt{K}-1)\sin\varphi\sin\psi}
\end{bmatrix}
\otimes
\begin{bmatrix}
1\\
e^{2\pi \js\frac{d?s}{\lambda}\sin\varphi\cos\psi}\\
\vdots\\
e^{2\pi \js\frac{d?s}{\lambda}(\sqrt{K}-1)\sin\varphi\cos\psi}
\end{bmatrix}.
\end{equation}
Here $d?s$ represents the inter-antenna spacing at the IRS whereas $\varphi$ and $\psi$ are the angles of elevation and azimuth, respectively. In simulations we set $d?b=d?s=\lambda/2$.
The angles $\psi_q$ and $\phi_q$ are uniformly distributed in the interval $[0,2\pi)$ and the $\varphi_q$'s are uniformly distributed in $[0,\pi)$. 
The channel of the direct link between the BS and each $m$-th single-antenna user, with $Q?{b-u}$ paths, is modeled as follows:
\begin{equation}
\label{eqn:chanl-b-u}
\hs_{\tn{b-u},m}~=~\sqrt{L(d_m)}\ds\sum_{q=1}^{Q?{b-u}}\cn_{m,q}\as?{BS}(\phi_{m,q}), \quad m=1, \cdots,M.
\end{equation}
Similar to the IRS-BS channel, $\cn_{m,q}\sim\CG\N(c_{m,q};0,1)$ and each angle $\phi_{m,q}$ is uniformly distributed in $[0,2\pi)$. The channel vectors in \eqref{eqn:chanl-b-u} are assumed to be independent across all users. Finally, the channel vector for the link between each $m$-th user, and the IRS with $Q?{s-u}$ channel paths, is modeled as follows:
\begin{equation}
\label{eqn:chanl-s-u}
\hs_{\tn{s-u},m}~=~\sqrt{L(d'_m)}\ds\sum_{q=1}^{Q?{s-u}}\cn_{m,q}\as?{IRS}(\varphi_{m,q},\psi_{m,q}), \quad m=1, \cdots, M.
\end{equation}
The term $L(d)=C?0(d/d?0)^{\eta}$ in \eqref{eqn:chanl-b-s}, \eqref{eqn:chanl-b-u}, \eqref{eqn:chanl-s-u} is the distance-dependent path-loss factor, where $C?0$ denotes the path-loss at a reference distance $d?0 = 1$ m, and $\eta$ is the path-loss exponent.
Moreover, to account for the line-of-sight (LOS) component, the gain of one channel path is set to at least of $5$ dB higher than the other path gains. \textcolor{dg}{To account for channel correlation effects, we have set the number of multi-path components lower than the number of BS antennas and the IRS antenna elements for the BS-IRS channel thereby making the channel matrix rank-deficient. Therefore, in simulations we have set the number of users lower than the rank of BS-IRS channel matrix $\h?{b-s}$.}
In the simulations, we fix $d?{IRS}=500$ m for the IRS-BS channel whereas the user-BS distance, $d$, and the user-IRS distance, $d'$, vary for each user according to its location from the BS and the IRS, respectively. In all simulations, we also set $C?0=-30 \ \tn{dB}$, $\eta=3.7$ (NLOS BS-user channel), $\eta=2.5$ (NLOS IRS-BS and IRS-user channels), $Q?{b-u}=2, \ Q?{s-u}=2, \ \epsilon=10^{-3}$ and $\sigma?w^2=-100$ dBm. The results are averaged over $1000$ independent Monte Carlo simulations.

The following two scenarios are studied. First, we consider the case where only the BS-IRS channel contains a LOS component. Then we consider the scenario where both the BS-IRS and the IRS-user channels have a LOS component but all the direct BS-user channels do not have a LOS component. The proposed VAMP-based algorithm is compared against the following four different configurations: 
\begin{itemize}
\item[i.] \textcolor{dg}{A MIMO system assisted by one IRS where the SDR technique is used to optimize the IRS reflection coefficients in combination with MMSE precoding.}
\item[ii.] A MIMO system assisted by one IRS where the joint optimization of the phase matrix the and the precoding is solved through alternate optimization and penalty-based ADMM. 
\item[iii.] A massive MIMO system with a large number of BS antennas with MMSE precoding. 
\item[iv.] An IRS-assisted MIMO system with unoptimized IRS phases and MMSE transmit \mbox{precoding.}
\end{itemize}
\vspace{-10pt}
\subsection{Benchmarking Metrics}

We use two metrics for performance evaluation, namely the \textit{sum-rate}, $\widehat{C}$, and the \textit{normalized root mean square error} (NRMSE) which are defined as follows:
\begin{equation}
\widehat{C}~=~\ds\sum_{m=1}^M\log_2\left(1+\frac{\abs*{\hs_m^\hr\fs_m}^2}{\sigma?w^2+\ds\sum_{i\neq m}\abs*{\hs_m^\hr\fs_i}^2}\right),
\end{equation}
where $\hs_m^H=\hs_{\tn{s-u},m}^\hr{\mathbf\Upsilon}\h?{b-s} +\hs_{\tn{b-u},m}^\hr$. 
\begin{equation}
\tn{NRMSE}(\alpha,{\mathbf\Upsilon},\f)~\triangleq~\frac{1}{\sqrt{M}}\sqrt{\norm{\alpha\left(\h?{s-u}^\hr{\mathbf\Upsilon}\h?{b-s}+\h?{b-u}^\hr\right)\f-\I_M}?F^2 + M\alpha^2\sigma?w^2}.
\end{equation}

\subsection{Performance Results With Perfect CSI}

\subsubsection{BS-IRS channel with LOS component}

\label{sec:sub-nlos-res}
\begin{figure}
\vspace{-20pt}
\centering
\begin{subfigure}{.333\textwidth}
\centering
\includegraphics[scale=0.4]{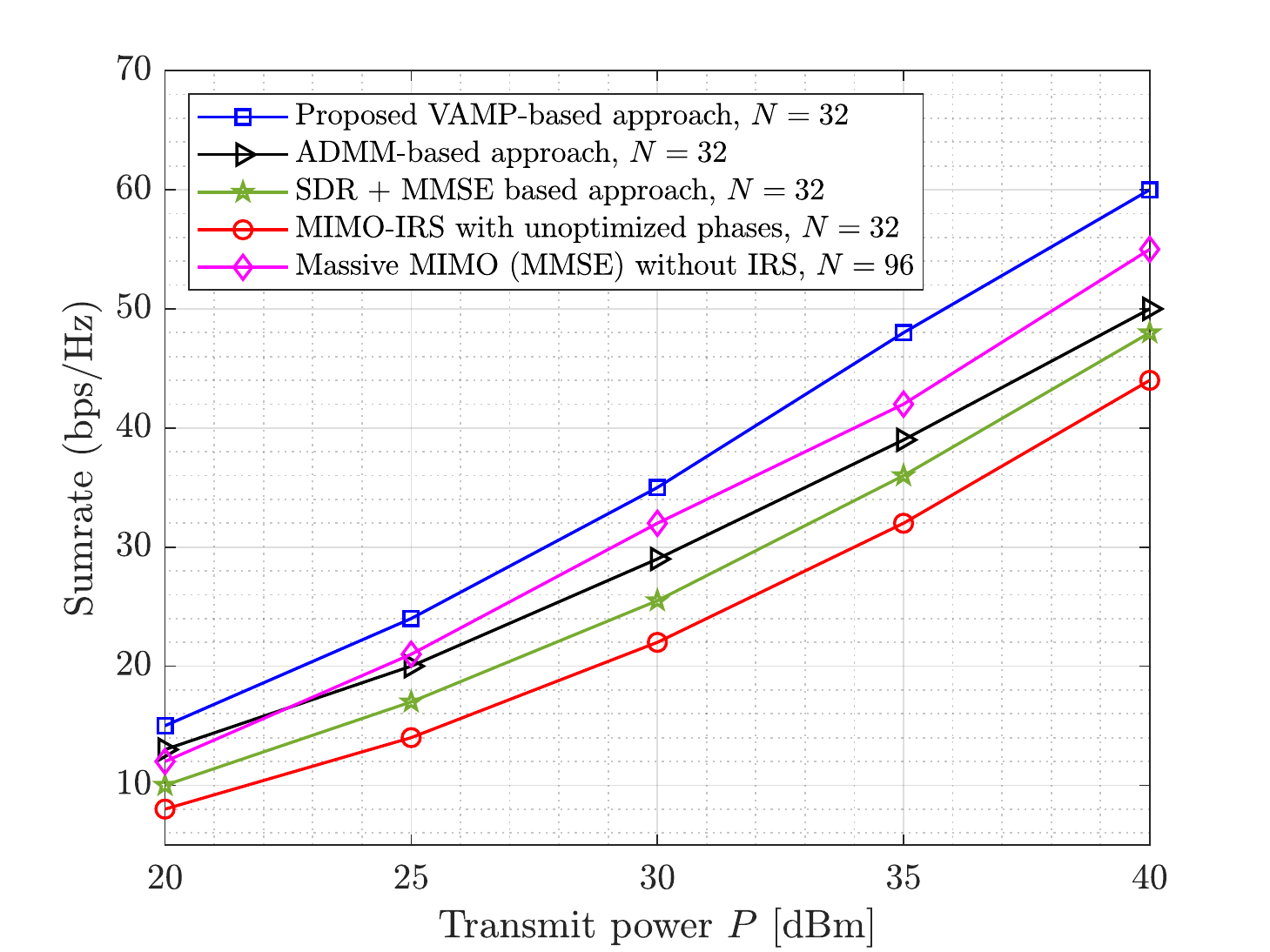}
\caption{\footnotesize $M=8$ and $K=256$.}
\label{fig:sum-p-nlos}
\end{subfigure}%
\begin{subfigure}{.334\textwidth}
\centering
\includegraphics[scale=0.4]{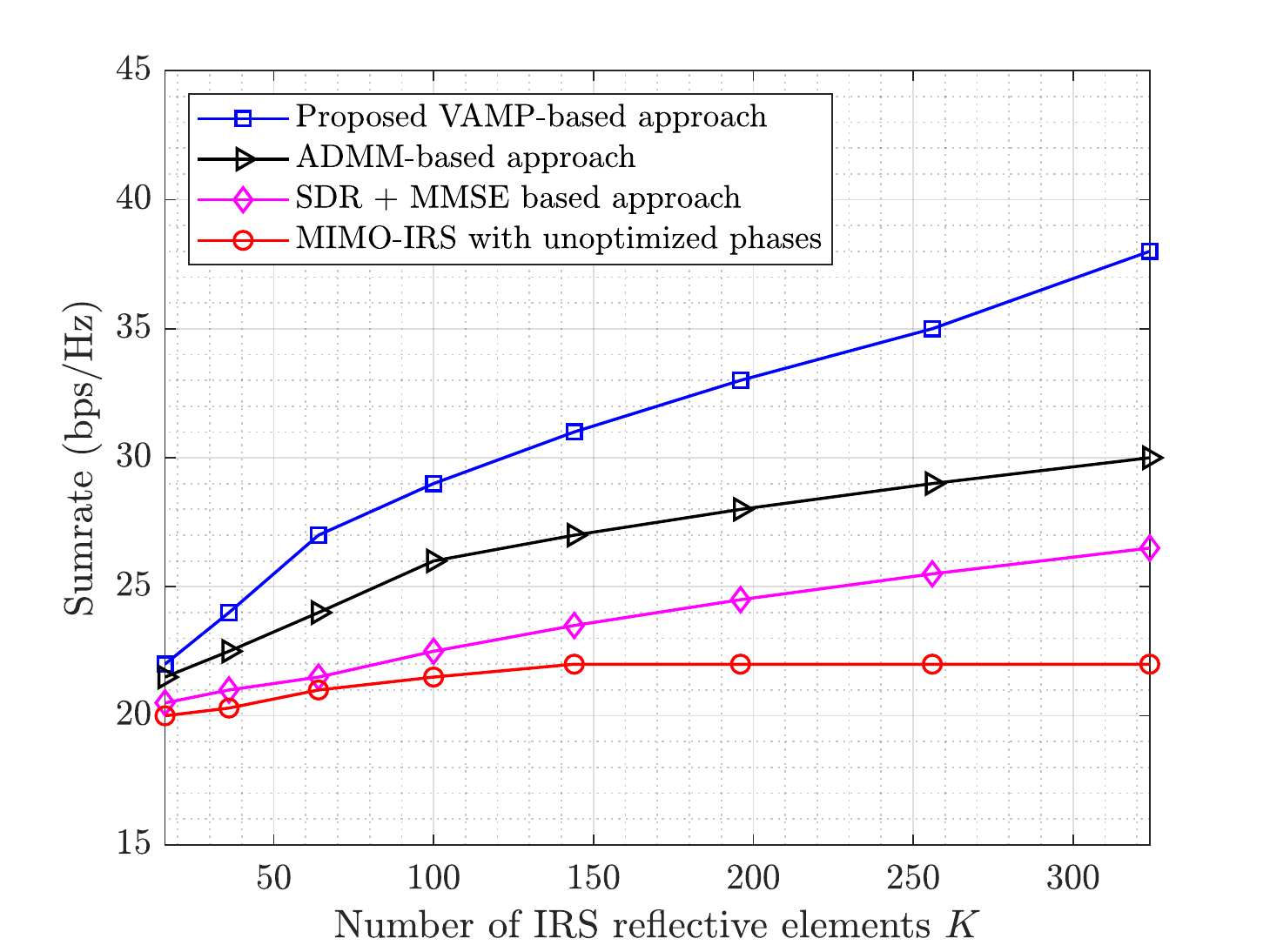}
\caption{\footnotesize $M=8, \ N=32$ and $P=30$ dBm.}
\label{fig:sum-elem-nlos}
\end{subfigure}%
\begin{subfigure}{.333\textwidth}
\centering
\includegraphics[scale=0.4]{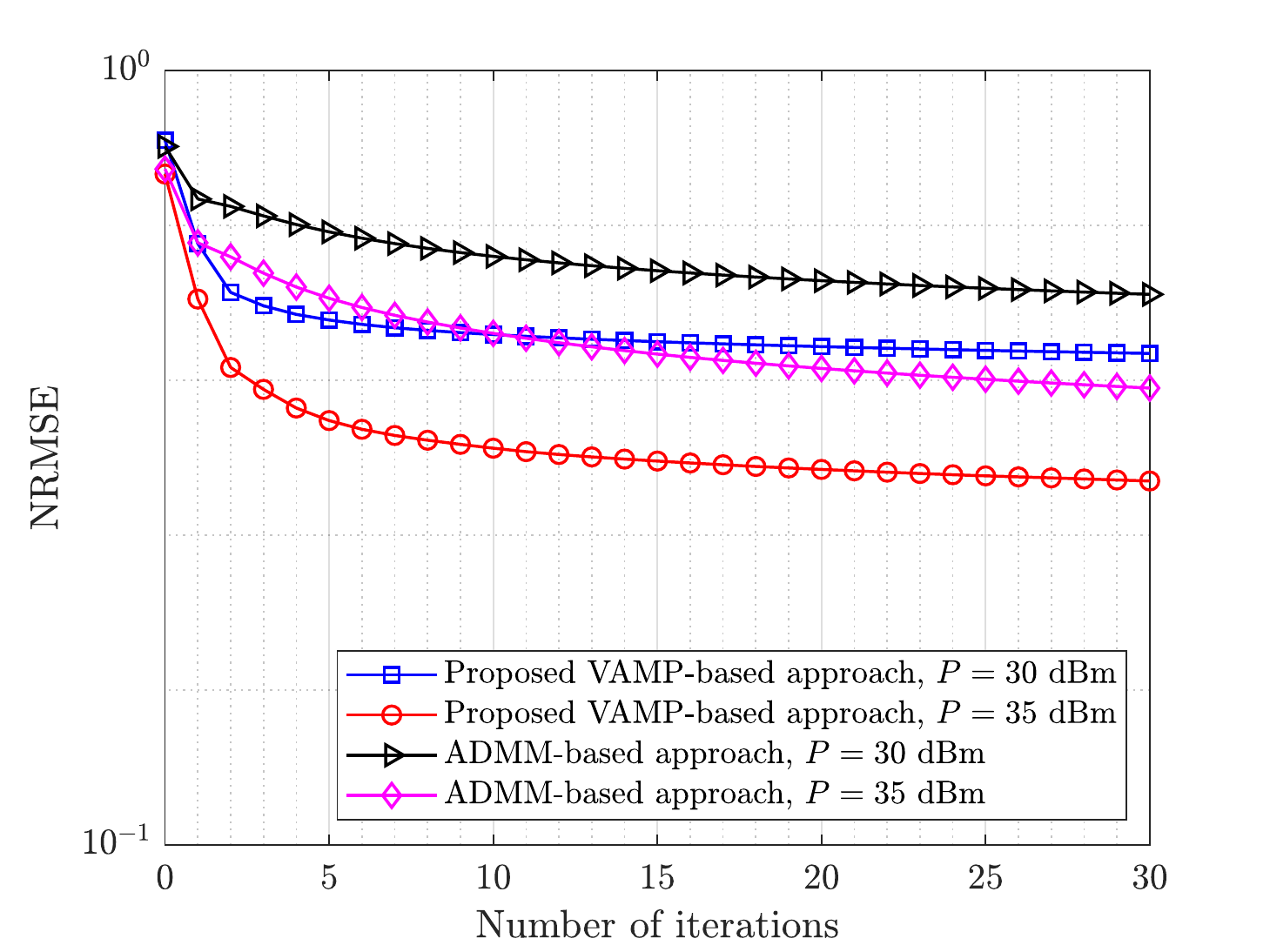}
\caption{\footnotesize $M=8, \ N=32$, and $K=256$.}
\label{fig:nrmse-nlos}
\end{subfigure}
\caption{LOS IRS-BS channel: (a) Sum-rate versus transmit power, (b) Sum-rate versus the number of IRS reflective elements, and (c) NRMSE versus the number of iterations (BS-user link excluded).}
\vspace{-25pt}
\end{figure}

This situation is encountered in a typical urban or suburban environments where the BS is located far away from the users and has no direct LOS component. However, the IRS is installed at a location where a LOS component is present in the BS-IRS link but not in the user-IRS link. Here we set the number of users to $M=8$ and the number of BS antennas to $N=32$ for every configuration except for massive MIMO for which we use $N=96$. 
Fig. \ref{fig:sum-p-nlos}, depicts the achievable sum-rate versus the transmit power, $P$, for the different considered transmission schemes. The proposed algorithm in this scenario outperforms the massive MIMO system even with a significantly smaller number of transmit antennas. 
VAMP automatically updates the stepsize at a per-iteration basis that leads to a faster convergence compared to other iterative algorithms
Since the proposed algorithm is based on VAMP, it outperforms the ADMM-based solution as it automatically updates the stepsize at a per-iteration basis (by means of calculating extrinsic information at each step) that leads to a faster convergence compared to ADMM, where the penalty parameter must be manually chosen. \textcolor{dg}{As per the IRS-assisted configuration, where one uses the SDR technique to optimize the IRS reflection coefficients, a significant gap is observed between the achieved sum-rates as compared to the proposed algorithm.}

Fig. \ref{fig:sum-elem-nlos} shows a plot of sum-rate against the number of IRS reflective elements. It is observed that even with a small number of active transmit antennas and merely ten paths between the IRS and the BS, the sum-rate for the proposed solution keeps increasing with the number of reflective elements. In contrast, the sum-rate saturates after a small gain when the IRS reflection coefficients are not optimized. \textcolor{dg}{Compared to the ADMM-based solution and the SDR method, the proposed algorithm shows higher throughput at every point.}

The convergence of the proposed algorithm is investigated in Fig. \ref{fig:nrmse-nlos} which depicts the NRMSE as a function of the number of iterations. Here we exclude the direct BS-user link to highlight the throughput of the BS-IRS-user link after optimizing the IRS phase shifts. Observe that the major portion of the gain is achieved in the first few iterations. The small number of iterations required for convergence in combination with the low per-iteration complexity makes the proposed algorithm very attractive from the practical implementation point of view. The superiority of the proposed VAMP-based algorithm over the ADMM-based approach stems from the feedback mechanism of VAMP. In fact, such feedback controls the weight given to the update of $\bm{\Upsilon}$ at each iteration compared to that of the preceding iteration. This is achieved with the help of scalar precision parameters that act as weighting coefficients for the $\bm{\Upsilon}$ updates that are computed in the current and the preceding iteration. In addition to the plots shifting downward, the increase in transmit power widens the gap between ADMM and the proposed VAMP-based algorithm. This demonstrates that the latter utilizes the available transmit power in a more efficient way than ADMM.

\begin{figure}
\vspace{-20pt}
\centering
\begin{subfigure}[b]{.333\textwidth}
\centering
\includegraphics[scale=0.4]{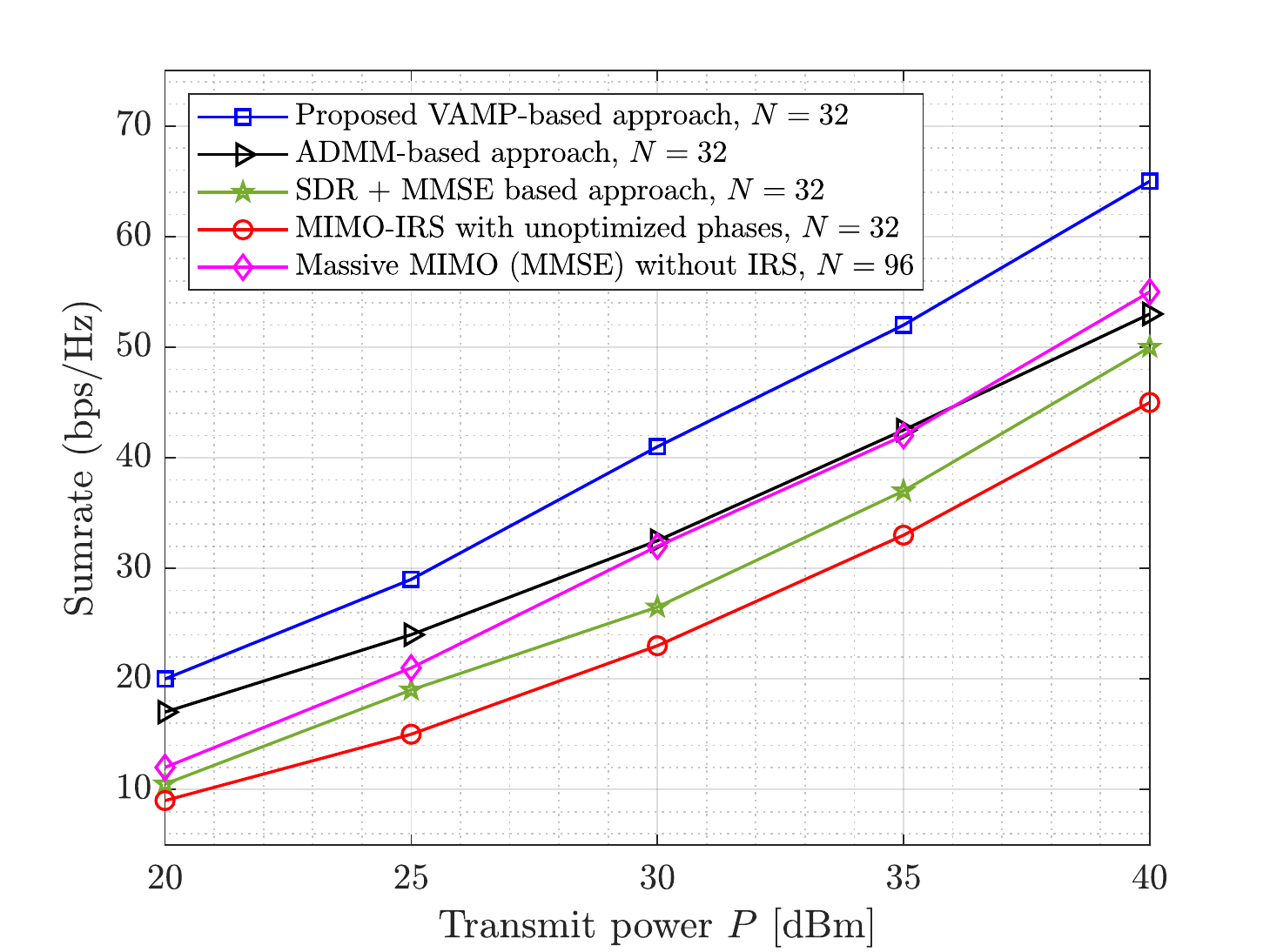}
\caption{\footnotesize $M=8$ and $K=256$.}
\label{fig:sum-p-los}
\end{subfigure}%
\begin{subfigure}[b]{.334\textwidth}
\centering
\includegraphics[scale=0.4]{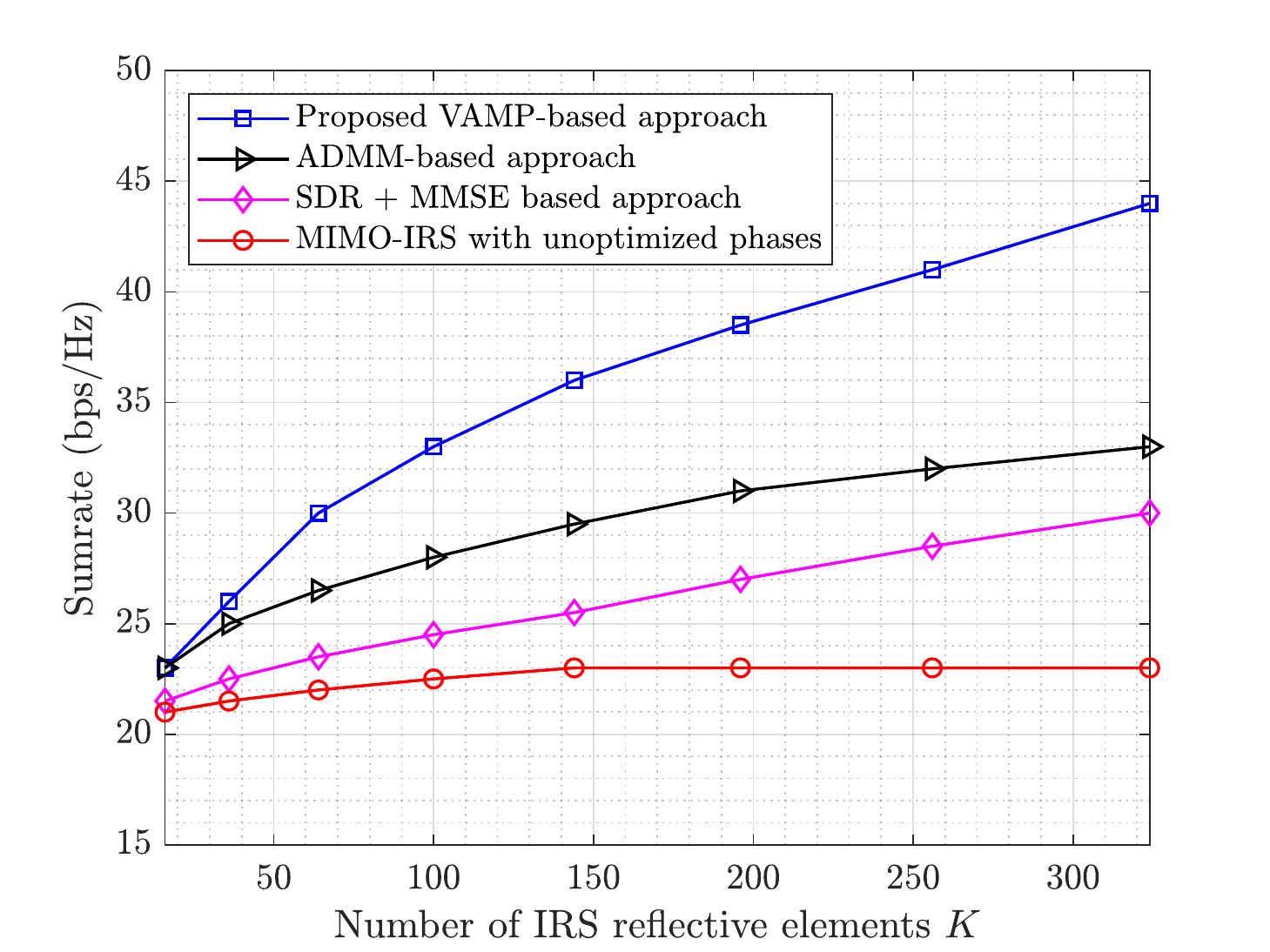}
\caption{\footnotesize $M=8, \ N=32$ and $P=30$ dBm.}
\label{fig:sum-elem-los}
\end{subfigure}%
\begin{subfigure}[b]{.333\textwidth}
\centering
\includegraphics[scale=0.4]{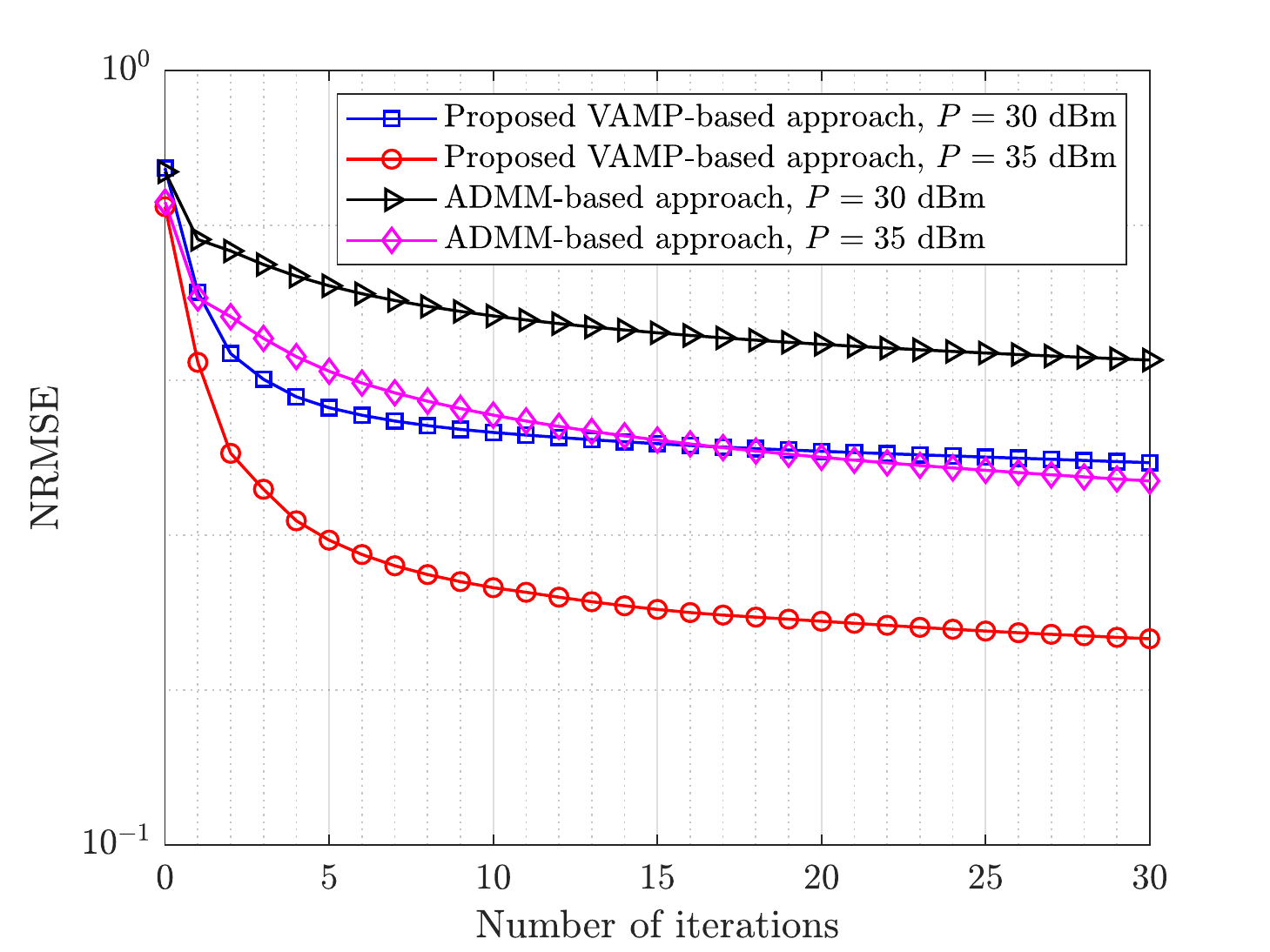}
\caption{\footnotesize $M=8, \ N=32$, and $K=256$.}
\label{fig:nrmse-los}
\end{subfigure}
\caption{LOS IRS-user and BS-IRS channels: (a) Sum-rate versus transmit power, (b) Sum-rate versus the number of IRS reflection elements, and (c) NRMSE versus the number of iterations (BS-user link excluded).}
\vspace{-20pt}
\end{figure}

\subsubsection{BS-IRS and IRS-user channels with LOS components}

\label{sec:sub-los-res}
Fig. \ref{fig:sum-p-los} illustrates the sum-rate versus the transmit power for this configuration. As expected, the results show that by adding a LOS component, the use of an IRS together with the proposed joint beamforming optimization solution yields considerably higher sum-rates compared to a massive MIMO system with no IRS. \textcolor{dg}{Moreover, although the ADMM-based solution now matches the performance of massive MIMO, the advantage of the proposed VAMP-based solution over the ADMM- and SDR-based solutions is higher when compared to the NLOS configuration.}

The results in Fig. \ref{fig:sum-elem-los}, i.e., sum-rate vs the number of IRS reflective elements, also exhibit the same trends as in the NLOS scenario yet with a broader gap between the curves, thereby, corroborating the superiority of the proposed solution. Intuitively, the presence of a LOS component helps the VAMP-based joint beamforming scheme to focus most of the transmit/reflected energy in that direction. This is clearly depicted in Fig. \ref{fig:nrmse-los}, where the NRMSE achieved by the proposed algorithm is approaching the NRMSE achieved by the ADMM-based solution but at almost $5$ dB lower transmit power.
\subsubsection{Practical IRS phase shifts}

\begin{figure}
\vspace{-10pt}
\centering
\begin{subfigure}{.333\textwidth}
\centering
\includegraphics[scale=0.4]{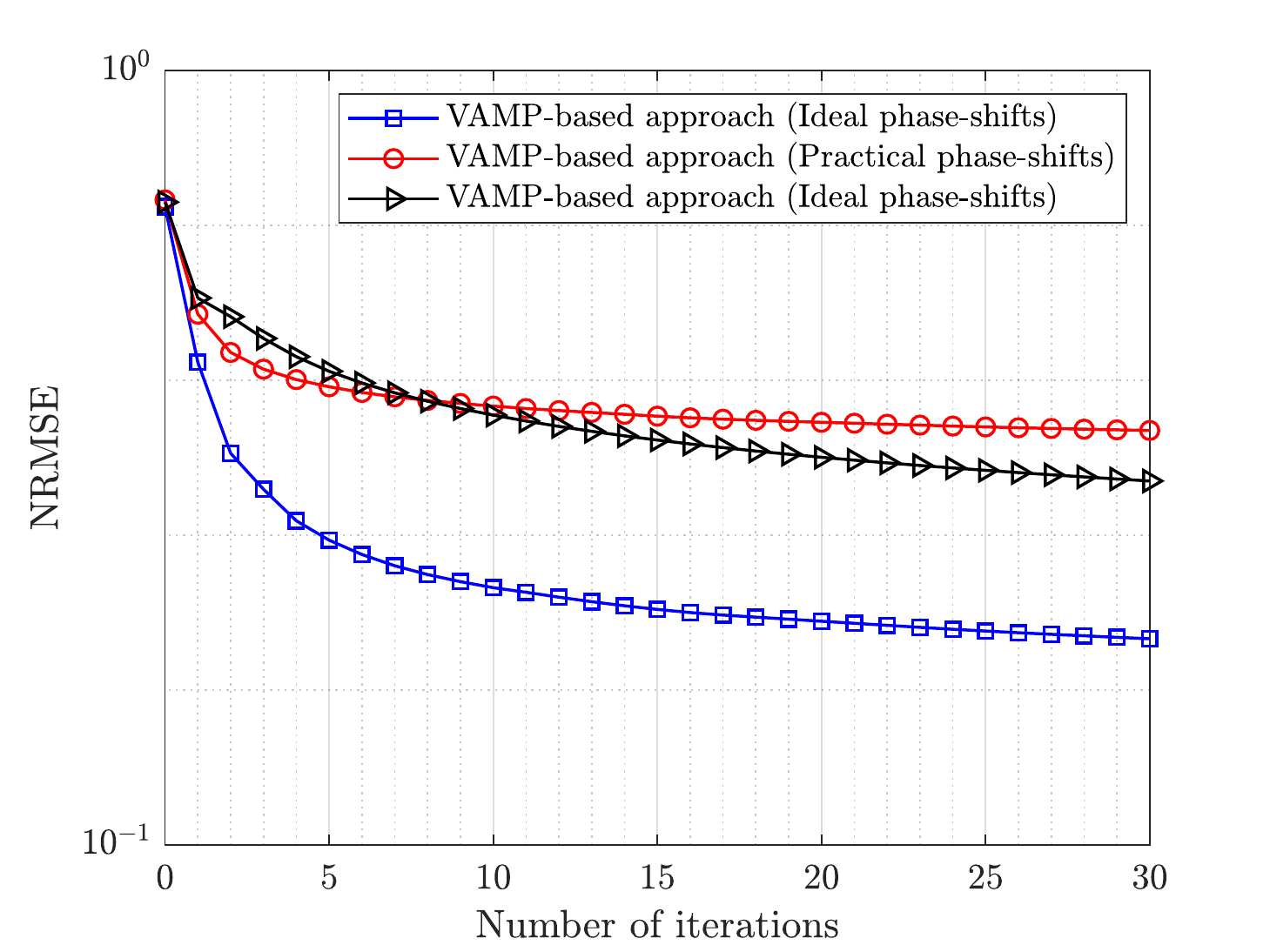}
\subcaption{\footnotesize $M=8, \ N=32$, $K=256$\\ and $P=30$ dBm.}
\label{fig:nrmse-los-react}
\end{subfigure}%
\begin{subfigure}{.334\textwidth}
\centering
\includegraphics[scale=0.4]{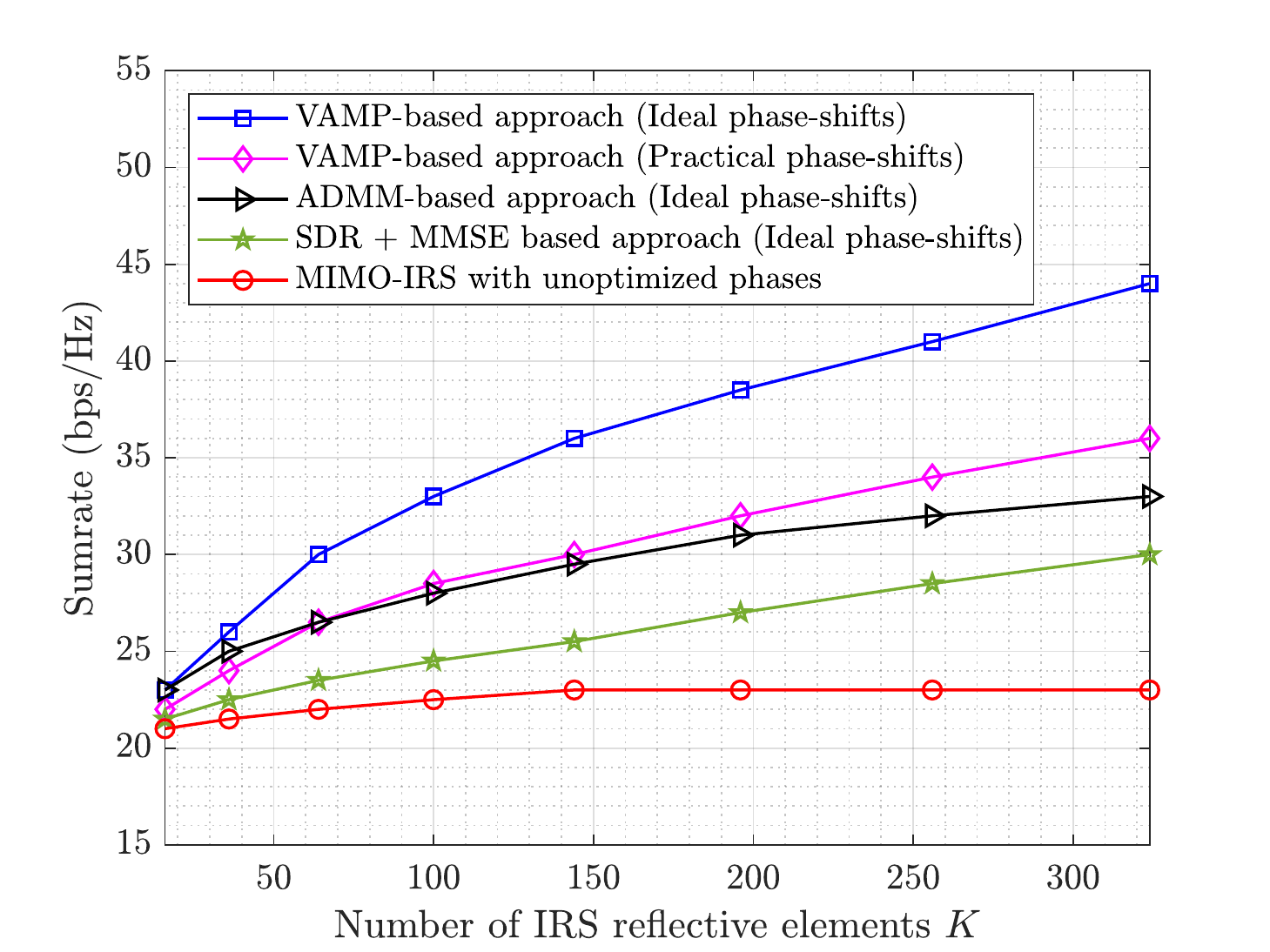}
\subcaption{\footnotesize $M=8, \ N=32$\\ and $P=30$ dBm.}
\label{fig:sum-elem-los-react}
\end{subfigure}%
\begin{subfigure}{.333\textwidth}
\centering
\includegraphics[scale=0.4]{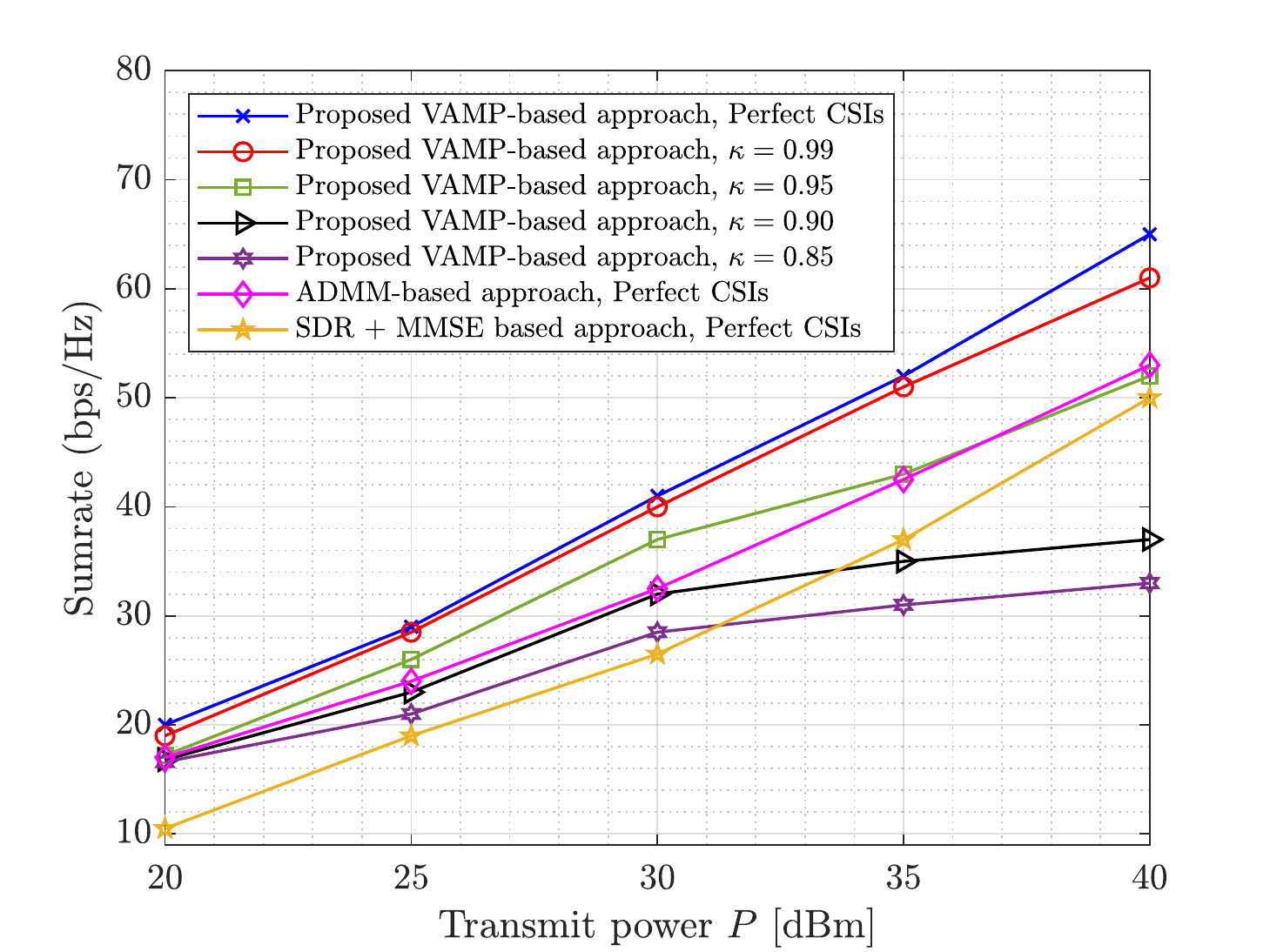}
\subcaption{\footnotesize $M=8, \ N=32$,\\ and $K=256$.}
\label{fig:imperfectcsi}
\end{subfigure}%
\caption{LOS IRS-user and BS-IRS channels: (a) NRMSE versus the number of iterations (BS-user link excluded), (b) Sum-rate versus IRS elements with practical phase shifts, and (c) Sum-rate versus transmit power under imperfect CSI.}
\vspace{-20pt}
\end{figure}

In this subsection, we assess the effect of replacing the unimodular constraint on the reflection coefficients by reactively loaded omni-directional elements. We use the same channel configuration as in Section \ref{sec:sub-los-res}. But, we rely on optimizing just the reactive part of the reflection coefficients. Therefore, as portrayed by Fig. \ref{fig:sum-elem-los-react}, the new constraint decreases the throughput when compared with the ideal phase shifters setup. However, the resulting sum-rate is still much higher than the one obtained by using unoptimized IRS reflection coefficients. \textcolor{dg}{In fact, when the number of IRS elements is higher than a certain value, the proposed approach with practical phase shifts achieves higher throughput than both the SDR- and ADMM-based solutions with ideal phase shifts.} Similarly, due to having less room for optimizing the reflection coefficients, Fig. \ref{fig:nrmse-los-react} shows that the NRMSE saturates sooner and at a higher value as compared to the case of a unimodular constraint (i.e., ideal phase shifts). Nonetheless, even with the more practical reactive load constraint, the resulting VAMP-based NRMSE is close to the NRMSE achieved by ADMM with ideal phase shifters.
\vspace{-10pt}
\textcolor{dg}{
\subsection{Performance Results With Imperfect CSI}
In this section, we measure the performance of the proposed solution in the presence of channel estimation errors. Specifically, we consider a scenario where pilot training followed by MMSE estimation algorithms are used to estimate the cascaded BS-IRS-users and the direct BS-user channels \cite{qauchannel1,qauchannel2}. We model the estimated channel matrix and vectors using the statistical CSI error model proposed in \cite{giffordcsi,annacsi,zhangcsi} as follows:
\begin{equation}
\widehat{\h}?{b-s}=\kappa\h?{b-s} + \sqrt{(1-\kappa^2)L(d?{IRS})}{\bm\Delta}?{b-s},
\end{equation}
\begin{equation}
\widehat{\hs}_{\tn{b-u},m}=\kappa\hs_{\tn{b-u},m} + \sqrt{(1-\kappa^2)L(d_m)}{\bm\delta}_{\tn{b-u},m}, \quad m=1, \cdots, M,
\end{equation}
\begin{equation}
\widehat{\hs}_{\tn{s-u},m}=\kappa\hs_{\tn{s-u},m} + \sqrt{(1-\kappa^2)L(d'_m)}{\bm\delta}_{\tn{s-u},m}, \quad m=1, \cdots, M,
\end{equation}
where $\kappa \in [0,1]$ denotes the channel estimation accuracy and ${\bm\Delta}?{b-s}$, ${\bm\delta}_{\tn{b-u},m}$ and ${\bm\delta}_{\tn{s-u},m}$ follow the circularly symmetric complex Gaussian (CSCG) distribution, i.e.,
\mbox{$\tn{vec}({\bm\Delta}?{b-s}) \sim \CG\N(\0, \1_{N \times N} \otimes \I_K)$,} ${\bm\delta}_{\tn{b-u},m} \sim \CG\N(\0, \I_N)$ and ${\bm\delta}_{\tn{s-u},m} \sim \CG\N(\0, \I_K)$. We first optimize the matrices $\f$ and ${\bm\Upsilon}$ under imperfect CSI and then use the exact CSI matrices to calculate the sum-rate.
Fig. \ref{fig:imperfectcsi} plots the sum-rate versus transmit power for different values of the channel estimation accuracy parameter $\kappa$. We also include plots for the other beamforming schemes under perfect CSI for reference. The results show the resilience of the proposed VAMP-based approach against small channel estimation errors. At low SNR, it is observed that the proposed design with a low channel estimation accuracy of $\kappa=0.85$ performs better than the SDR based approach and nearly as good as the ADMM-based approach under perfect CSIs. Moreover, the performance loss with a high channel estimation accuracy value of $\kappa=0.99$ is negligible.
}
\textcolor{dg}{
\section{Convergence, Optimality, and Complexity Analysis}
\label{sec:complexity-con-opt}
According to the monotone convergence theorem in real analysis \cite{monotone}, a monotonically decreasing sequence with a lower bound is convergent. In our case, since the objective function,
\begin{equation}
\norm{\alpha\h?{s-u}^\hr{\bm\Upsilon}\h?{b-s}\f-(\I_M-\alpha\h?{b-u}^\hr\f)}?F^2 + M\alpha^2\sigma?w^2
\end{equation}
has a lower bound of zero, the proposed algorithm will always converge to a solution if the MSE monotonically decreases in both steps of the algorithm, i.e., the step of optimizing the phase-shifts (VAMP part) and the step of optimizing the active precoding. For the latter, we have a closed-form optimal solution. Therefore, it is necessary that the MSE decreases monotonically inside the VAMP step in every iteration to guarantee the convergence of the entire algorithm. In practice, most of the approximate message passing-based algorithms (including VAMP) add damping steps inside the algorithm to avoid any oscillations in the resultant MSE and thus, ensuring convergence \cite{ranganvamp}.
The lines 18 and 20 inside the VAMP part of the \textbf{Algorithm 4} are, respectively, replaced by the damped versions:
\begin{equation}
\widetilde{\gamma}_t=\varrho\gamma_{t-1}\left\langle\dk_t\right\rangle/\left(\frac{\mathsmaller K}{\mathsmaller R?{\tiny BA}} -\left\langle\dk_t\right\rangle\right) +(1-\varrho)\widetilde{\gamma}_{t-1}.
\end{equation}
\begin{equation}
\widehat{\bm\upsilon}_t=\varrho\gs_1(\widetilde{\rs}_t)+(1-\varrho)\widehat{{\bm\upsilon}}_{t-1},
\end{equation}
for all iterations $t>1$ where $\varrho \in (0,1]$ is a suitably chosen damping factor.}
\begin{table}[!ht]
\centering
\caption{\textcolor{dg}{Comparison between the CPU execution time of the proposed VAMP-based algorithm, the ADMM-based algorithm and the SDR-based algorithm for different design configurations. The algorithms terminate when $\abs*{\tn{NRMSE}_t-\tn{NRMSE}_{t-1}}<10^{-3}\tn{NRMSE}_{t-1}$ or $t>100$.}}
\textcolor{dg}{
\label{table:process-time}
\begin{tabular}{|p{0.3\textwidth}|P{0.14\textwidth}|P{0.14\textwidth}|P{0.14\textwidth}|P{0.14\textwidth}|}
\hline 
\vspace{5pt}\center{\textbf{Design Parameters}}  & \textbf{VAMP-based algorithm} $\BO(MN(K+N))$ (msec)  & \textbf{ADMM-based algorithm} $\BO(MN(K+N))$ (msec)  & \textbf{SDR-based algorithm} $\BO(MN + K^6)$ (msec)\\
\hline 
\hline
 $M=2, \ N=16, \hspace{7pt} K=64$ & $14$ & $26$ & $2100$\\
\hline 
\hline
 $M=4, \ N=32, \hspace{7pt} K=256$ & $104$ & $340$ & $12500$\\
\hline 
\end{tabular}
}
\end{table}
\textcolor{dg}{
The optimality of the proposed VAMP-based approach can be investigated through statistical state evolution analysis of the proposed algorithm which we have left for a future work. Please note that for non convex optimization problems like optimizing the phase-shifts matrix under uni-modular constraint, asymptotic (for large matrix sizes) optimality can be claimed for i.i.d. matrices, if the proximal functions (projector function and its derivative inside \textbf{Algorithm 4}) are shown to be Lipschitz continuous, and the state evolution analysis reveals that the VAMP-based algorithm has only one fixed point \cite{ranganvamp,barbieroptimal}.
For implementation purpose, we choose the maximum possible value for precision tolerance, $\epsilon$, for which the mean square error (MSE) approximately saturates before the algorithm is stopped. For the proposed solution we have found out that $\epsilon=10^{-3}$ does the trick as the MSE achieved by choosing any lower values for $\epsilon$ is approximately equal to the MSE achieved by choosing $\epsilon=10^{-3}$. The maximum number of iterations, $T?{max}$, is a hardware-dependent parameter and is manually chosen to have a limit on the number of iterations.}

Note that, by utilizing the Kronecker structure, we avoid any large matrix multiplication or even taking SVD of Kronecker or Khatri-Rao product of matrices.
Let $\A=\alpha\h?{s-u}^\hr$ and $\B=(\h?{b-u}\f)^\tr$. For our system model, the matrices $\A$ and $\B$ are of the same size $M \times K$. Assuming that the matrices $\A$ and $\B$ are of full rank, the complexity of the truncated SVDs of the matrices is of $\BO(M^2K)$. \textcolor{dg}{The computational complexity of the column-wise Khatri-Rao product in line $10$ and the following operations in lines $11$ and $12$ of \textbf{Algorithm \ref{algo:overall}} has a complexity of $\BO(M^2K)$.} The Kronecker product of two vectors in line $13$ and the component-wise operations of vectors in lines $16$ and $17$ are of order $\BO(M^2)$. The projector function and its derivative has a complexity in the order of $\BO(K)$. \textcolor{dg}{The functions $g_3\left(\h\right)$ and $g_4\left(\h\right)$ can be implemented efficiently by using the matrix inversion lemma, thereby entailing a complexity of $\BO(M^3+MN^2)$. The complexity of all other matrix multiplications elsewhere including the LMMSE part is of order $\BO(MNK+M^2K)$. Therefore, the overall per-iteration complexity of the algorithm is of order $\BO(M^3+M^2K +MNK +MN^2)$. Since $M<N$ and $M<K$ in our case, the overall per-iteration complexity simplifies to \mbox{$\BO(MN(K+N))$ or $\BO(MNK)$ for $N\leq K$.}}

Table \ref{table:process-time} provides a comparison of CPU (central processing unit) run time between the VAMP-based approach, the ADMM-based approach  and the SDR-based approach for different design configurations. For comparison purpose, we measure the time until the NRMSE saturates with a tolerance, $\epsilon$ , therefore we run the algorithms until \mbox{$\abs*{\tn{NRMSE}_t-\tn{NRMSE}_{t-1}}<\epsilon\tn{NRMSE}_{t-1}$} or $t>T?{MAX}$, while setting $\epsilon=10^{-3}$ and $T?{MAX}=100$. We set the channel simulation parameters as in Section \ref{sec:sub-los-res} with $P=30$ dBm. The algorithms are simulated on MATLAB R2020a on a laptop PC having a Core i7-4720HQ processor and 16 GB of RAM with Windows 10 operating system. As expected, the simulation results confirm that the proposed approach is significantly faster in terms of \mbox{convergence time, especially when there is a high number of IRS antenna elements.}
\vspace{-30pt}
\section{Conclusion}
\textcolor{dg}{
\label{sec:conclusion}
We investigated the problem of joint active and passive beamforming design for an IRS-assisted downlink multi-user MIMO system under both ideal and practical models for the IRS phase shifts. The associated joint non-convex optimization has been formulated under sum-MMSE criterion. Using alternating minimization, the joint optimization has been decomposed into two sub-optimization tasks, i.e., optimizing the IRS phase shifts and the BS precoding matrix separately. Regarding the phase shifts, we have presented a novel approach that relies on the approximate message passing framework to solve the associated sub-optimization problem. For this, we have first extended the traditional VAMP algorithm, and then used the extended version to find a local optimum, for the phase-shifts matrix under both ideal and practical constraints. The optimal precoder at the BS, however, was found in closed-form using Lagrange optimization. Simulation results illustrate the superiority of the proposed approach over existing beamforming schemes (e.g., the SDR-and ADMM-based approaches) both in terms of throughput and convergence speed. The results also illustrate that the reduction in the throughput of the system under more restrictive phase shifts is not significant. Moreover, it has been shown that the performance of the proposed approach is largely unaffected by small channel estimation errors.
Its optimality can also be investigated through state evolution analysis which is left for a follow-up work.
Since the proposed solution provides flexibility in terms of choosing the constraint on the IRS reflection coefficients, it opens up the possibility of solving the joint optimization problem using more \mbox{physically consistent models for the IRS reflection elements.}
}
\vspace{-10pt}
\appendices
\section{}

\label{apd:A}
\noindent
We solve the following optimization problem:
\begin{equation}
\ds\argmin_{\chi} \ f(\chi),
\end{equation}
where
\begin{equation}
f(\chi)~\triangleq~ \abs*{\widetilde{r}+\frac{1}{1+\js\chi}}^2,
\end{equation}
in which $\chi\in\R$ and $\widetilde{r} \in \C$. Expanding the objective function, we re-express it as follows:
\begin{equation}
\label{eqn:apda-obj}
\ds\argmin_{\chi} \ \widetilde{r}^*\widetilde{r} +\frac{\widetilde{r}^*}{1+\js\chi}+\frac{\widetilde{r}}{1-\js\chi} +\frac{1}{(1-\js\chi)(1+\js\chi)}.
\end{equation}
Let $a\triangleq\Re\left\lbrace\widetilde{r}\right\rbrace$ and $b\triangleq\Im\left\lbrace\widetilde{r}\right\rbrace$. We substitute $a$ and $b$ into \eqref{eqn:apda-obj} and simplify the objective function as follows:
\begin{equation}
\label{eqn:apda-obj-expl}
\ds\argmin_{\chi} \ a^2+b^2 + \frac{1+2a}{1+\chi^2} - \frac{2b\chi}{1+\chi^2}.
\end{equation}
By defining $c\triangleq(1+2a)$, we take the derivative w.r.t. $\chi$ and set it to zero to obtain:
\begin{equation}
\label{eqn:apda-first-der}
f'(\chi)~=~-\frac{2b(1-\chi^2)}{(1+\chi^2)^2}-\frac{2c\chi}{(1+\chi^2)^2}~=~0.
\end{equation}
Simplifying \eqref{eqn:apda-first-der} leads to:
\begin{equation}
\label{eqn:apda-quad}
b\chi^2-c\chi-b~=~0.
\end{equation}
The roots of the quadratic equation in \eqref{eqn:apda-quad} are real and distinct and are given by:
\begin{equation}
\chi_1~=~\frac{c+\sqrt{c^2+4b^2}}{2b},
\end{equation}
and
\begin{equation}
\chi_2=\frac{c-\sqrt{c^2+4b^2}}{2b}.
\end{equation}
where $b\neq0$. By taking the second derivative of the objective function in \eqref{eqn:apda-obj-expl} w.r.t. $\chi$ and resorting to some straightforward algebraic manipulations, we also obtain:
\begin{equation}
\label{eqn:apda-second-der}
f''(\chi)~=~\frac{2}{(1+\chi^2)^3}(6b\chi-2b\chi^3 + 3c\chi^2-c).
\end{equation}
Substituting  $\chi=\chi_1$ in \eqref{eqn:apda-second-der} and simplifying the result yields:
\begin{equation}
f''(\chi_1)~=~\frac{1}{(1+\chi_1^2)^3}\left(\frac{1}{b^2}\left(c^3+c^2\sqrt{c^2+4b^2}\right)+4\left(c+\sqrt{c^2+4b^2}\right)\right).
\end{equation}
Since $b\neq0$, we have $c^2\sqrt{c^2+4b^2}>\abs*{c^3}$ and $\sqrt{c^2+4b^2}>\abs*{c}$ which implies that $f''(\chi_1)>0$. Similarly, we have:
\begin{equation}
f''(\chi_2)~=~\frac{1}{(1+\chi_2^2)^3}\left(\frac{1}{b^2}\left(c^3-c^2\sqrt{c^2+4b^2}\right)+4\left(c-\sqrt{c^2+4b^2}\right)\right)<0, \quad b\neq0.
\end{equation}
Thus, we choose:
\begin{equation}
\chi\<{opt}~=~\chi_1~=~\frac{1+2a+\sqrt{(1+2a)^2+4b^2}}{2b},
\end{equation}
Interestingly, the solution $\chi_1$ results in the same sign for both $\Im\left\lbrace-(1+\js\chi_1)^{-1}\right\rbrace$ and $\Im\left\lbrace\widetilde{r}\right\rbrace$.
\bibliographystyle{IEEEtran}
\bibliography{IEEEabrv,references} 
\end{document}